\documentclass[a4paper,fleqn,usenatbib,onecolumn]{mnras}

\usepackage[T1]{fontenc}
\usepackage{ae,aecompl}

\usepackage{graphicx}	
\usepackage{amsmath}	
\usepackage{amssymb}	
\usepackage{multirow,bigstrut}
\usepackage{geometry}
\usepackage{slashbox}

\title[The formation of Galactic BH LMXBs]
{On the Formation of Galactic Black Hole Low-Mass X-ray Binaries}

\author[Wang et al.]{
Chen Wang,$^{1,2}$ Kun Jia$^{1,2}$ and
Xiang-Dong Li$^{1,2}$\thanks{E-mail: lixd@nju.edu.cn}
\\
$^{1}$Department of Astronomy, Nanjing University, Nanjing
210046, China\\
$^{2}$Key laboratory of Modern Astronomy and Astrophysics
(Nanjing University), Ministry of Education, Nanjing 210046, China\\
}

\date{Accepted XXX. Received YYY; in original form ZZZ}

\pubyear{2015}

\begin{document}
\label{firstpage}
\pagerange{\pageref{firstpage}--\pageref{lastpage}}
\maketitle

\begin{abstract}
Currently, there are 24 black hole (BH) X-ray binary systems that
have been dynamically confirmed in the Galaxy. Most of them are
low-mass X-ray binaries (LMXBs) comprised of a stellar-mass BH and a
low-mass donor star. Although the formation of these systems has
been extensively investigated, some crucial issues remain
unresolved. The most noticeable one is that, the low-mass companion
has difficulties in ejecting the tightly bound envelope of the
massive primary during the spiral-in process. While initially
intermediate-mass binaries are more likely to survive the common
envelope (CE) evolution, the resultant BH LMXBs mismatch the
observations. In this paper, we use both stellar evolution and binary population synthesis
to study the evolutionary history of BH LMXBs. We test
various assumptions and prescriptions for the supernova mechanisms
that produce BHs, the binding energy parameter, the CE efficiency,
and the initial mass distributions of the companion stars. We obtain
the birthrate and the distributions of the donor mass, effective
temperature and orbital period for the BH LMXBs in each case. By
comparing the calculated results with the observations, we put
useful constraints on the aforementioned parameters. In particular, we
show that it is possible to form BH LMXBs with the standard CE
scenario if most BHs are born  through failed supernovae.
\end{abstract}

\begin{keywords}
binaries: general $-$ black hole physics $-$ X-ray:
binaries $-$ stars: evolution
\end{keywords}



\section{Introduction}

There may be around $10^8-10^9$ stellar-mass black holes (BHs) in the
Galaxy \citep[e.g.,][]{Brown1994,Timmes1996}, but only two dozen of
them have been dynamically confirmed in X-ray binary (XRB) systems,
in which a BH is accreting from its companion star \citep[][for
recent reviews]{RemillardMcClintock2006,CasaresJonker2014}. A large
fraction of the BH XRBs are close binaries with orbital periods
$P_{\rm orb}< 1$ day. The spectral types of the companion stars
range from A2V to M1V, indicating that most of them are less massive
than $2\, M_{\odot}$, so these XRBs are correspondingly classified as
low-mass X-ray binaries (LMXBs). All known BH LMXBs are transient
systems, spending most of their time in quiescence with X-ray
luminosities below $\sim \rm 10^{32}\, erg\,s^{-1}$, and occasionally
exhibiting outbursts when the X-ray luminosities can rise up to $\sim
10^{37}-10^{39}\, \rm erg\,s^{-1}$.

Galactic BH LMXBs are believed to evolve from primordial binaries
consisting of a massive star and a low- or intermediate-mass
companion \citep[][for a review]{Li2015}.
The standard formation scenario involves a common envelope (CE)
phase, during which the binary's orbital separation reduces
drastically due to the loss of the orbital kinetic energy
\citep{Paczynski1976, van den Heuvel1983}, since the orbital
separations of the present day binaries are orders-of-magnitude
smaller than the radii that the massive primaries can reach during
their evolution. When the primary star evolves and overflows its
Roche lobe, mass transfer proceeds on a dynamical timescale because
the primary is much more massive than the secondary. The secondary
cannot accrete all the mass transferred onto it, and an envelope
forms enshrouding the two stars. Then the secondary star spirals
into the envelope, and the orbital energy is used to expel the envelope
\citep[for reviews of
CE evolution, see][]{TaamSandquist2000,Ivanova2013}. If surviving the CE event, the
binary can continue to evolve; otherwise the two stars would merge
to be a single star.  Though the CE stage is too short-lived to be
detected, indirect observational evidence has emerged
\citep[e.g.,][]{DrakeSarna2003, Sion2012,Ivanova2013}.

BHs are thought to be the evolutionary outcomes of sufficiently massive stars
\citep[$\gtrsim 20-25\, M_{\odot}$; e.g.,][]{WW1995,fry01}. However, the
envelope of such stars are generally too tightly bound for a
low-mass secondary to expel \citep{Podsiadlowski2003}, thus a merger
is more likely to take place, unless the primary has lost a
significant fraction of its envelope before the CE phase through
stellar winds \citep{Wiktorowicz2013}, or abnormally high values for
the CE ejection efficiency are adopted
\citep{KielHurley2006,YungelsonLasota2008}.

An alternative scenario is that the secondaries in BH LMXBs are
initially intermediate-mass stars, which can eject the primary's
envelope more easily. Observational evidence for this scenario comes
from the CNO-processed material in XTE J1118$+$480 from its
ultraviolet spectra \citep{Haswell2002}. It is, however, difficult
for an intermediate-mass X-ray binary (IMXB) to evolve to
short-period systems, unless some unusual angular momentum loss
mechanisms are invoked \citep[e.g.,][]{Justham2006,ChenLi2006}.
Meanwhile, these binaries reveal a discrepancy between the
calculated effective temperatures and the observed spectral types of
the donor stars. The reason lies in that nuclear burning in the
center of an intermediate-mass star causes a higher effective
temperature than an ordinary low-mass star, though it has lost a
significant fraction of its mass during the evolution
\citep{Justham2006}.

Clues to the initial mass range for the donors in BH LMXBs can be
derived from the spin evolution of the BHs, since the BH spins
increase with accretion \citep{Thorne1974}. Two methods, i.e., X-ray
continuum-fitting \citep{Zhang1997,Davis2005} and modeling
relativistic reflection \citep{Fabian1989} have been employed to
estimate the spins of about 20 BHs  \citep[][and references
therein]{McClintock2014, Fabian2014}. Assuming that BHs in LMXBs are
born with negligible spins, \cite{Fragos2015} showed that the
initial masses of the secondaries in  short-period binaries are
generally $\lesssim 1-2\,M_{\sun}$.

From another point of view,  approaches have been taken in the study
of CE evolution by refining the energy budget for the ejection as
well as the core-envelope binding energy. For example,
\cite{Ivanova2002} argued that nuclear fusion should be added into
the energy sources that may contribute to the envelope ejection.
\cite{Podsiadlowski2010} proposed that a thermonuclear runaway which
is caused by the mixing hydrogen from the secondary into the
helium-burning shell of the primary would assist the ejection. When
treating the binding energy of the envelope,
\cite{IvanovaChaichenets2011} pointed out that there might be mass
outflows during the slow spiral-in stage, when the orbital energy is
balanced with the enthalpy rather than merely the internal energy of
the envelope. That treatment can lower the binding energy by a
factor of $\sim 2-5$, and help allow for a low-mass companion to
survive the CE phase. It is also noted that the value of the binding
energy is sensitive to the definition of the boundary between the
remnant core and the ejected envelope \citep[e.g.,][]{Tauris2001}.
\cite{Ivanova2011} advocated that, the He core of a giant would
experience a thermal readjustment stage after the CE event. Thus the
boundary between the remnant core and the ejected envelope can be
determined as the place where the hydrogen burning shell had maximal
compression prior to CE evolution.

Another issue in the formation of BH LMXBs is the mass range for the
BH progenitors. Current supernova (SN) theories cannot reliably
predict which stars form BHs instead of neutron stars (NSs),
although they are generally thought to be more massive than $\sim
20-25\ M_{\odot}$. The BH masses in the 24 XRBs have been
dynamically measured to be in the range of $\sim 2.7\, M_{\odot}$ to
$\gtrsim 15\, M_{\odot}$ \citep[][and references
therein]{CasaresJonker2014}. However, there is statistical evidence
for  the presence of a dearth of NSs or BHs with masses $\sim 2-5
M_{\odot}$ \citep{bai98,Ozel2010,Ozel2012,Farr2011,kre12,Kiz2013}.
This is in contrast with the traditional thought that the
distribution of BH masses should decay with mass
\citep[e.g.,][]{fry99,fry01}, suggesting that the physics of SN
explosions that lead to the formation of BHs is still unclear
\citep{Fryer2012,bel12}. Recently, \cite{Kochanek2014} argued that
zero-age main sequence stars with mass $\gtrsim \rm 17\  M_{\odot}$
might experience an unsuccessful explosion and evolve to BHs, while
stars with mass around $8-17\ M_{\odot}$ experience a successful SN
explosion and become NSs. Though this bifurcation is not well
understood and may be related to the compactness parameter that
describes the density profile outside the iron core
\citep{OConnorOtt2011, Clausen2015, Kochanek2014}, the failed SN
mechanism can naturally account for the observed $\sim 2-
5\,M_{\sun}$ gap between NSs and BHs. \cite{Horiuchi2014}
proposed that this mechanism can interpret the red supergiant
problem and the SN rate problem, which concerns the absence of red
supergiants with masses $\sim 16-30\ M_{\odot}$ as the progenitors
of Type IIP SNe, and the deficit of the observed cosmic SN rate
compared to the observed cosmic star formation rate, respectively.

In this paper we investigate the formation of Galactic BH LMXBs
incorporating the influence of the aforementioned factors with both
stellar evolution and
binary population synthesis (BPS) methods. In section 2, we calculate
and compare the binding energy parameters in various models, which
are then used in our BPS study. In section 3, we introduce the
assumptions and prescriptions adopted in our BPS calculations. The
results in various kinds of models are presented  and
compared with observations in section 4.
We summarize the results and discuss their implications in section 5.
\section{The binding energy parameter}

In the standard CE model the change in
the binary's orbital energy $\Delta E_{\rm orb}$
is responsible for the ejection of the CE, which is described by the
following equations \citep{Webbink1984},
 \begin{equation}\label{equation1}
   E_{\rm bind}=\alpha_{\rm CE}\Delta E_{\rm orb},
 \end{equation}
and
  \begin{equation}\label{equation2}
   \Delta E_{\rm orb}=\frac{GM_{\rm core}M_2 }{2a_{\rm f} }
   -\frac{GM_1 M_2}{2a_{\rm i}},
 \end{equation}
where $E_{\rm bind}$ is the binding energy of the envelope,
$\alpha_{\rm CE}$ the efficiency of converting the orbital energy to
kinetic energy to eject the CE, $G$ the gravitational constant,
$M_1$ the primary's mass and $M_{\rm core }$ its core mass, $M_2$
the mass of the secondary, and $a_{\rm i}$ and $a_{\rm f}$  the
orbital separations just before and after the CE phase,
respectively.

When only gravitational binding energy is considered,  $E_{\rm
bind}$ is given by
 \begin{equation}\label{equation3}
    E_{\rm bind}= \int_{M_{\rm core}}^{M_1}-\frac{GM(r)}{r}dm.
 \end{equation}
 If the internal energy of the stellar matter (including both thermal energy
 and recombination energy) also  contributes to the binding energy, then we have
 \begin{equation}\label{equation4}
    E_{\rm bind}= \int_{M_{\rm core}}^{M_1}\left[-\frac{GM(r)}{r}+U\right ] dm,
 \end{equation}
where $U$ is  internal energy \citep{Han1994}. Recently,
\cite{IvanovaChaichenets2011} proposed a modified energy criterion
taking into account the mass outflows during the spiral-in stage. In
this so-called enthalpy model, the binding energy is expressed as
\begin{equation}\label{equation5}
    E_{\rm bind}= \int_{M_{\rm core}}^{M_1}\left[-\frac{GM(r)}{r}+U+\frac{P}{\rho}\right] dm,
\end{equation}
where $P$ and $\rho$ are the pressure and the density of the gas,
respectively. Since the $P/\rho$ term is non-negative, the
absolute value of $E_{\rm bind}$ becomes smaller in this case.

For convenience, \cite{deKool1990} introduced a parameter $\lambda$
characterizing the central concentration of the donor's envelope to
describe the binding energy:
\begin{equation}\label{equation6}
    E_{\rm bind} = -\frac {GM_1 M_{\rm env}}{\lambda a_{\rm i} r_{\rm L}},
\end{equation}
where $M_{\rm env}$ is the mass of the donor star's  envelope,  and
$r_{\rm L}= R_{\rm L}/ a_{\rm i}$ is the ratio of the Roche lobe
radius and the orbital separation at the onset of CE. Generally,
once a star fills its Roche lobe, $a_{\rm i}r_{\rm L}$ is taken to
be the stellar radius.

Combining Eqs.~(\ref{equation3})--(\ref{equation6}), one can obtain different $\lambda$s from
the gravitational binding energy, the total energy and the enthalpy
prescriptions, namely $\lambda_{\rm g}$, $\lambda_{\rm b}$ and
$\lambda_{\rm h}$, respectively. We calculate their values using an
updated version (7624) of the Modules for Experiments in Stellar Astrophysics (MESA) \citep{Paxton2011,Paxton2013,Paxton2015} evolution code. We consider stars with initial
masses in the range of $19-60\, M_{\odot}$, which securely covers the
range of the stellar-mass BH progenitors (more massive stars produce
negligible BHs according to the initial mass function and will evolve to Wolf-Rayet (WR)
stars with almost all the hydrogen envelopes lost due to strong stellar winds). The chemical
composition of the stars is taken to be $X=0.7$ and $Z=0.02$. The
mixing length parameter $\alpha = l/H_{\rm p}$ is set to be 2.0. For massive stars, stellar winds
may have non-ignorable influence on the $\lambda$
parameter \citep{Podsiadlowski2003}, but their loss rates have not been accurately determined.
Here, we adopt two prescriptions for the wind loss rate. The first one is taken from
\cite{Hurley2000} and \cite{Vink2001} for O and B stars in different
stages, denoted as Wind1 hereafter, while the other one, denoted as Wind2, adopts the maximum value of the
prescriptions above in all the evolutionary stages, to be consistent with
\cite{XuLi2010a,XuLi2010b}, and to set an upper limit for the influence of the stellar wind.
The effect of stellar rotation is ignored, because in our calculation for LMXB formation,
CE evolution usually occurs during Case C mass transfer when the primary star has evolved
to be a supergiant star with very slow rotation due to stellar evolution and/or tidal synchronization.

\begin{figure}
\centerline{\includegraphics[width=1.00\textwidth]{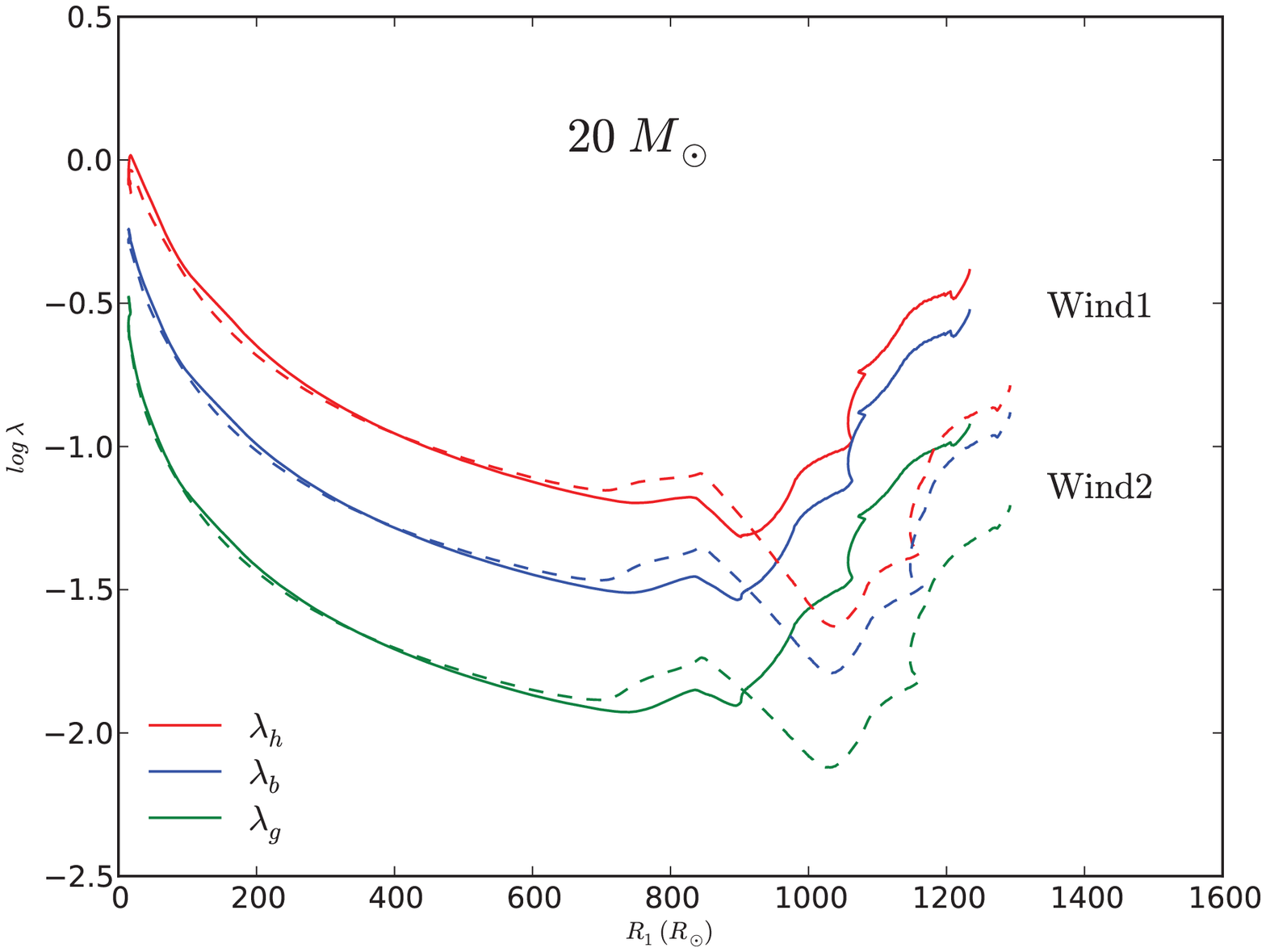}}
\caption{Evolution of the three types of binding energy parameters $\lambda$
with the stellar radius $R$ for a $\rm Pop.\ I$ star with initial
mass of $20\ M_{\sun}$. The the red, blue and green lines represent $\lambda_{\rm h}$,
$\lambda_{\rm b}$, and $\lambda_{\rm g}$, and the solid and dashed lines
represent the results with the Wind1 and Wind2 prescriptions, respectively.\label{figure1}}
\end{figure}

\begin{figure}
\centerline{\includegraphics[width=1.00\textwidth]{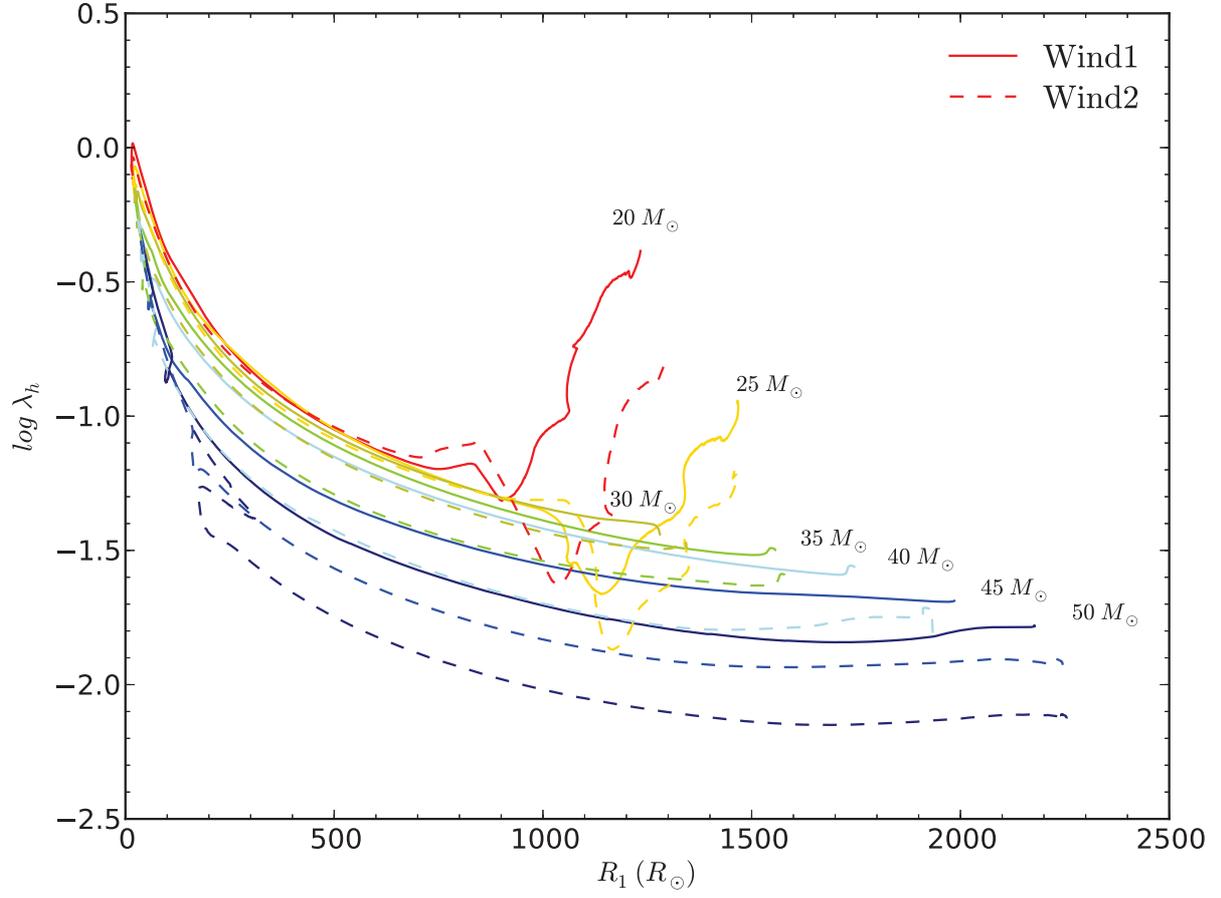}}
\caption{The $\lambda_{\rm h}$ values as a function of the stellar
radius for stars with different mass. The solid and dashed lines represent the results with
the Wind1 and Wind2 prescriptions, respectively. \label{figure2}}
\end{figure}

\begin{figure}
\centerline{\includegraphics[width=1.00\textwidth]{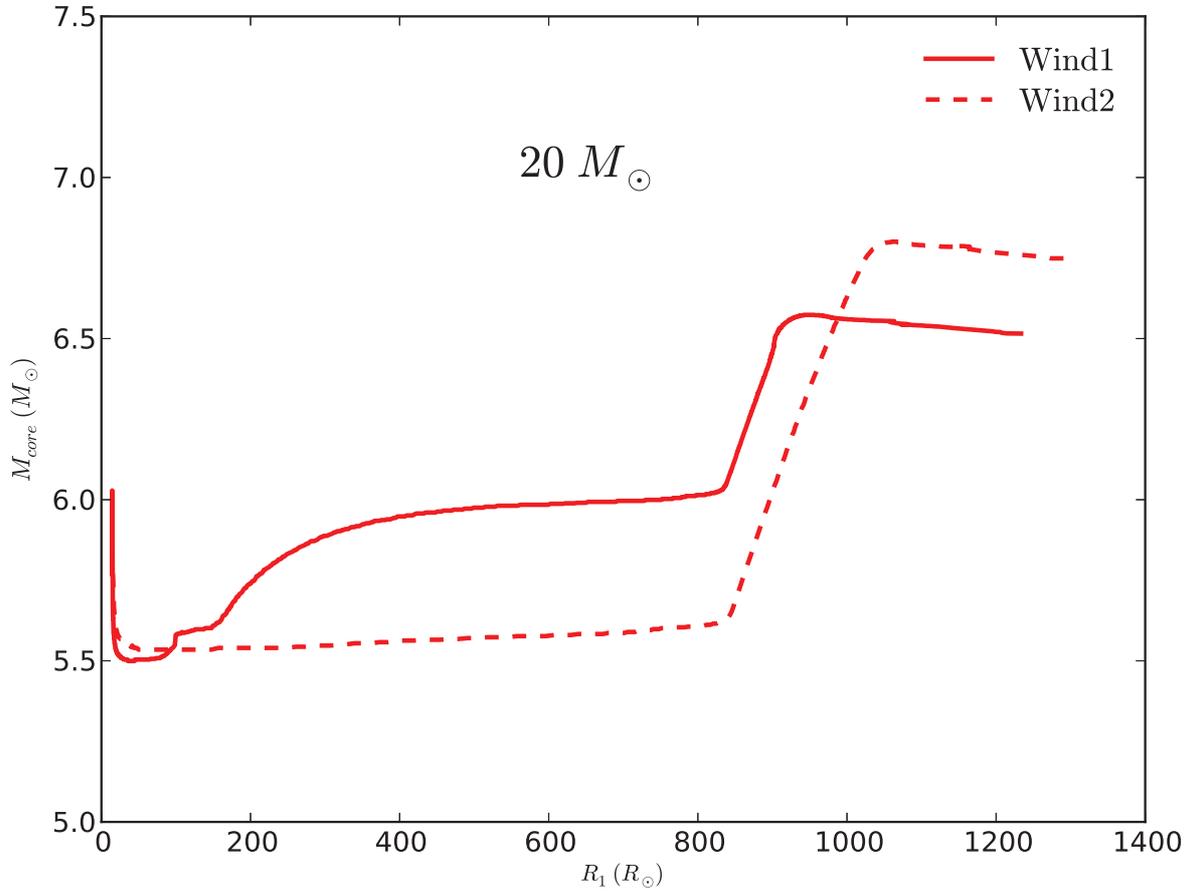}}
\caption{The core mass as a function of the stellar radius
 for a $20\, M_{\sun}$ star. Here, the boundary of the core and the envelope is
 determined to be the maximal compression point in the hydrogen burning shell.
 The solid and dashed lines represent the results with the Wind1 and Wind2 prescriptions,
 respectively. \label{figure3}}
\end{figure}

Our calculations show that  the evolution of $\lambda_{\rm h}$, $\lambda_{\rm b}$, and
$\lambda_{\rm g}$ generally have the similar trend in most of the
evolutionary stages, i.e., decreasing with increasing stellar
radius. However, for stars less massive than $30\, M_{\sun}$,
$\lambda$s increase again when they have ascended the asymptotic giant branch and
developed a deep envelope \citep[see also][]{Podsiadlowski2003}.
As an illustration, Fig.~\ref{figure1} shows different $\lambda$ values as a
function of the stellar radius $R$ for a $20\,M_{\odot}$ star,
and Fig.~\ref{figure2} shows the $\lambda_{\rm h}$ values as a function of the
stellar radius for stars with different masses. The solid and dashed
lines correspond to the Wind1 and Wind2 prescriptions, respectively.
In Fig.~\ref{figure1} the red, blue, and green lines describe the evolution of
$\lambda_{\rm h}$, $\lambda_{\rm b}$, and $\lambda_{\rm g}$
respectively in each case. It is seen that the values of $\lambda$ with
the Wind1 prescription are about three times as large as those with the Wind2 prescription
(the binding energy with the Wind1 prescription is half of the value with the Wind2 prescription
considering different envelope masses), indicating that
the binding energy parameter for massive stars
is sensitive to wind loss \citep{Podsiadlowski2003}.
We plot the core mass against the stellar radius in Fig.~\ref{figure3},
and list the stellar parameters along the evolutionary tracks in Table~\ref{table1} in the two cases.

We also use an updated stellar evolution code called EV originally developed by
\cite{Eggleton1971,Eggleton1972} to calculate the values of $\lambda$s, and
find that they are  several times smaller than those with
the MESA code in the Wind2 case \citep[see also][]{XuLi2010a,XuLi2010b}.
The reason for this difference is that
stars are generally more compact (especially for stellar structure within the hydrogen
burning shell) during late evolution when modeled with the EV code.  We choose to use
the results with the MESA code in the following because it adopts much denser grids for
stellar structure than the EV code.

\cite{Justham2006} pointed out that a low-mass
($< 1\,M_{\odot}$) star  is unable to eject the BH progenitor's envelope
unless $\lambda\gtrsim 0.15$. In our calculation, this condition could be achieved for
both $\lambda_{\rm h}$ and $\lambda_{\rm b}$ if the masses of the  BH progenitors
are $<25\ M_{\odot}$.  Binaries with more massive
progenitor stars are unable to survive CE evolution even if $\lambda_{\rm h}$
is adopted (see also Fig.~\ref{figure2}).

\section{Binary population synthesis}
We use an updated version of the BPS code developed by
\cite{Hurley2002} to study the formation of BH LMXBs. The code
originates from the rapid single-star evolution code written by
\cite{Hurley2000}. It involves various kinds of processes such as
mass transfer, accretion, CE evolution, SN kicks, tidal
interactions, and angular momentum loss. We have made quite a few
modifications of the BPS code to follow the formation and evolution
of XRBs \citep{Shao2014}. We briefly describe some considerations
adopted in this work.

\subsection{CE evolution}
Given the binding energy parameter $\lambda$, we can estimate the
post-CE separation with the following equation:
\begin{equation}\label{equation7}
    \frac{a_{\rm f}}{a_{\rm i}}= \frac{M_{\rm core} M_{\rm 2}}{M_1}
    \frac{1}{M_{\rm 2}+2M_{\rm env}/\alpha_{\rm CE} \lambda r_{\rm L}}.
\end{equation}
If neither the He core nor the secondary star fills its Roche lobe,
the binary is regarded to survive the CE phase and can continue to
evolve.

Generally the value of $\alpha_{\rm CE}$ is less than unity, and there
is indication that the CE efficiency could be low \citep{Ohlmann2016}.
Moreover, it may vary with stellar structure and
the binary parameters. Indeed, it has been suggested that
$\alpha_{\rm CE}$ may depend on the component mass and the orbital period in the studies of
the formation of post-CE white dwarf binaries
\citep[e.g.,][]{DeMarco2011,Davis2012}.
However, it is not known whether this can be applied to more massive binaries,
so we still adopt constant values (1 and 0.2) for $\alpha_{\rm CE}$. It is noted that
$\alpha_{\rm CE}$ and $\lambda$ are usually combined to affect CE evolution.

\subsection{Remnant mass}
Besides CE evolution, the SN explosions play an important role
in  the formation of BH LMXBs. The SN mechanisms are relevant to
both the formation rate of XRBs and the mass
distribution of the remnant BHs.

When determining the BH masses $M_{\rm BH}$, one needs to account
for the measured masses of known BHs and the $\sim 2-5\,M_{\sun}$
gap between the NS and BH masses without any fine-tuning of stellar
mass loss. We consider both the prescriptions based on the
neutrino-driven convection-enhanced SN mechanism, which is called
the rapid SN mechanism, and the failed SN mechanism. In the former
case the gravitational remnant mass is described by
\citep{Fryer2012},
\begin{equation}\label{equation8}
    M_{\rm rem}=0.9 M_{\rm rem,bar}=0.9(M_{\rm proto}+M_{\rm fb}).
\end{equation}
Here $M_{\rm proto}=1.0\,M_{\odot}$ is the mass of the proto-compact
object and
\begin{equation}\label{equation9}
   \left \{
     \begin{array}{lcl}
      M_{\rm fb}=0.2 \ M_{\odot}    &  &   M_{\rm CO}  < 2.5\,M_{\odot} \\
      M_{\rm fb}=0.286M_{\rm CO}-0.514 \ M_{\odot}  &     & 2.5\,M_{\odot} \leq  \textit{M}_{\rm CO} < 6.0 \,M_{\odot} \\
      M_{\rm fb}=f_{\rm fb}(M_1-M_{\rm proto})  &   & M_{\rm CO}  \geq 6\,M_{\odot},
    \end{array}
    \right.
\end{equation}
with
\begin{equation}\label{equation10}
   \left \{
     \begin{array}{lcl}
      f_{\rm fb}=1.0 &   &  {6.0\ M_{\odot} \leq  \textit{M}_{\rm CO} < 7.0 \ M_{\odot}} \\
      f_{\rm fb}=a_1M_{\rm CO}+b_1 &  &   {7.0\ M_{\odot} \leq  \textit{M}_{\rm CO} < 11.0 \ M_{\odot}} \\
      f_{\rm fb}=1.0  &   & {M_{\rm CO}  \geq 11.0\ M_{\odot} },
    \end{array}
    \right.
\end{equation}
where $M_{\rm CO}$ is the CO core mass, $a_1=0.25-1.275/(M_1-M_{\rm
proto})$, and $b_1=-11a_1+1$.

The latter prescription is under the assumption that the BH formation is
controlled by the compactness of the stellar core at the time of
collapse: low compactness stars are more likely to explode as SNe
and produce NSs, while high-compactness stars are more likely to
evolve to failed SNe that produce BHs \citep{OConnorOtt2011}. In
this case we assume $M_{\rm BH}=M_{\rm He}$  or $M_{\rm CO}$, where
$M_{\rm He}$ is the He core masses prior to core collapse
\citep[e.g.,][]{Smith2011,SmithArnett2014,ShiodeQuataert2014,Sukhbold2014,
Clausen2015,Kochanek2014,Kochanek2015}.
The value of $M_{\rm He}$ is determined as
follows \citep[][for a detailed description]{Hurley2000}. First, the core mass $M_{\rm c}$ at
the end of the Hertzsprung-gap is set according to the initial mass of the star
\citep[see Eq.~28 of][]{Hurley2000}. Then the evolution of the core mass
during the giant branch is determined by combining the power-law
core mass-luminosity relation
\begin{equation}\label{equation11}
   L=\min(BM^{q}_{\rm c},DM^{p}_{\rm c}),
\end{equation}
where $p=5$, $q=2$, and $B$ and $D$ depend on the stellar mass,
and the energy conservation equation for hydrogen burning
\begin{equation}\label{equation12}
   L=EX_{\rm e}\dot{M}_{\rm c},
\end{equation}
where $X_{\rm e}$ is the envelope mass fraction of hydrogen and $E$ is the
specific energy release. Thus
\begin{equation}\label{equation13}
  \dot{M}_{\rm c}=\min(A_{\rm H}BM_{\rm c}^{q},A_{\rm H}DM_{\rm c}^{p}),
\end{equation}
where $A_{\rm H}=1/EX_{\rm e}$ represents hydrogen rate constant. A similar procedure is adopted
to calculate the CO core mass, with $A_{\rm H}$ replaced by
$A_{\rm He}=7.66\times 10^{-5}M_{\sun}L_{\sun}^{-1}$ Myr$^{-1}$.

In all of the cases we require
that $M_{\rm BH}\geq 3M_{\odot}$. We do not consider the formation
of BHs by accretion-induced collapse of NSs in XRBs.
We also assume that a natal ``kick" is imparted on the newborn BHs,
similar as NSs. The kick velocity is set to be inversely
proportional to the remnant mass, i.e., $V_{\rm kick}({\rm
BH})=(3M_{\sun}/ M_{\rm BH}) V_{\rm kick}(\rm NS)$, where $V_{\rm
kick}(\rm NS)$ is the kick velocity for NSs, which follows a
Maxwellian distribution with $\sigma = \rm 265\ km\ s^{-1}$
\citep{Hobbs2005}.


\subsection{Initial parameters}

We consider the BH LMXBs evolved from primordial binaries with both
incipient low-mass and intermediate-mass secondaries. We assume a
constant star formation rate of $5\,M_{\sun}$yr$^{-1}$ \citep{s78},
and evolve $10^7$ primordial binaries with the primary mass ($M_1$)
ranging from $19\,M_{\odot}$ to $60\,M_{\odot}$ following the
\citet{Kroupa1993} initial mass function, and the secondary mass
($M_2$) from $0.1\,M_{\odot}$ to $6\,M_{\odot}$. For the initial
mass ratio $q=M_2/M_1$, we consider both flat distribution $n(q)=1$
with $q$ randomly distributed between 0 and 1, and a ``twin"-like
distribution with $n(q)\propto q$. The initial separation $a$ of the
binary components varies between $3\,R_{\odot}$ and
$10^4\,R_{\odot}$, with a flat distribution for $\ln a$.
All binaries are assumed to be initially in circular orbits.  We stop the calculation when
the donor mass in LMXBs drops down to $0.1\,M_{\odot}$, or the evolution time
exceeds 15 Gyr. We summarize the models and the prescriptions
employed in Table.~\ref{table2}.

\section{Results}

\subsection{The donor mass distribution}

\begin{figure}

     \begin{tabular}{cc}
     \begin{minipage}[h,t]{3.3in}
     \includegraphics[width=3.3in]{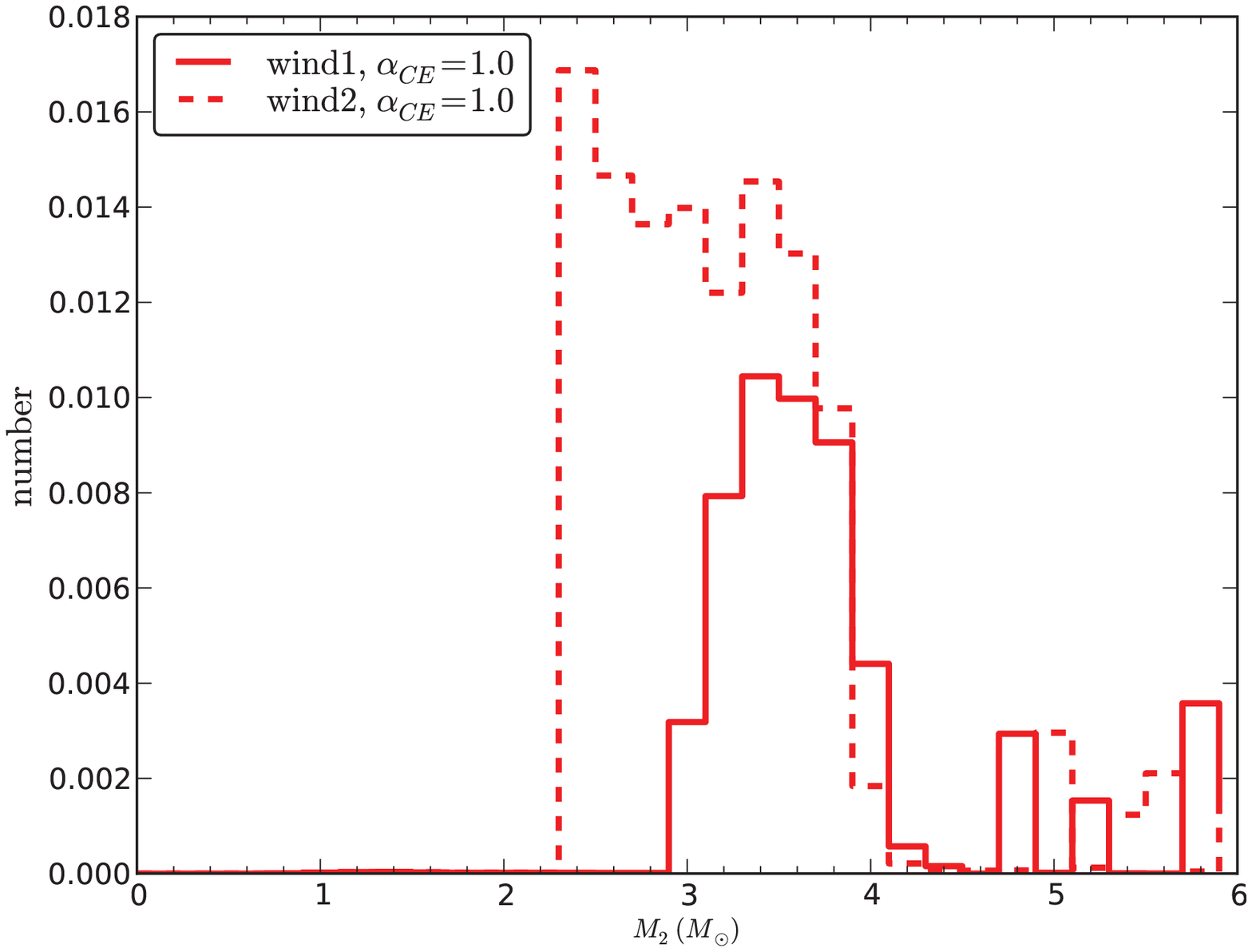}
     \end{minipage}

     \begin{minipage}[h,t]{3.3in}
     \includegraphics[width=3.3in]{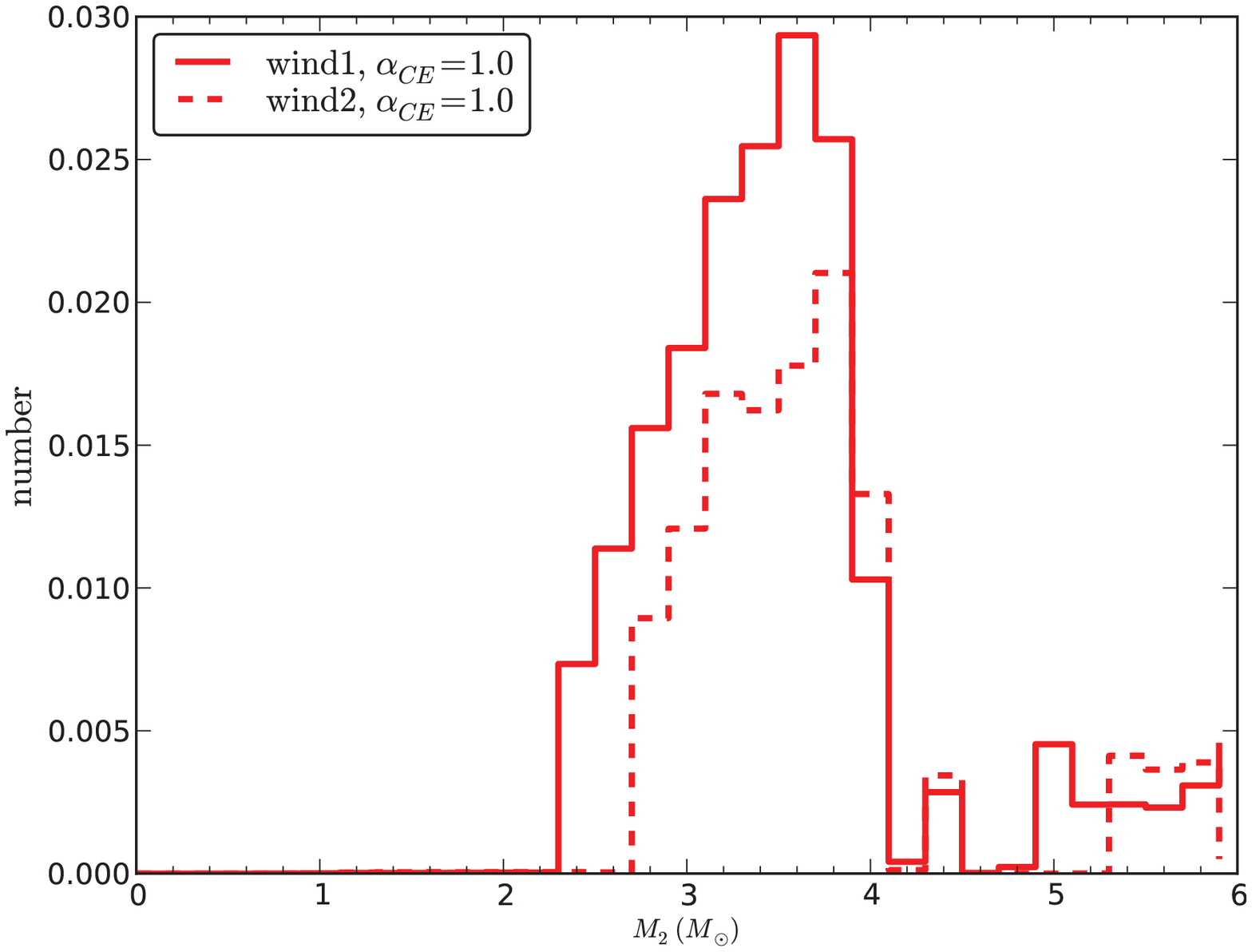}
     \end{minipage}\\

     \begin{minipage}[h,t]{3.3in}
     \includegraphics[width=3.3in]{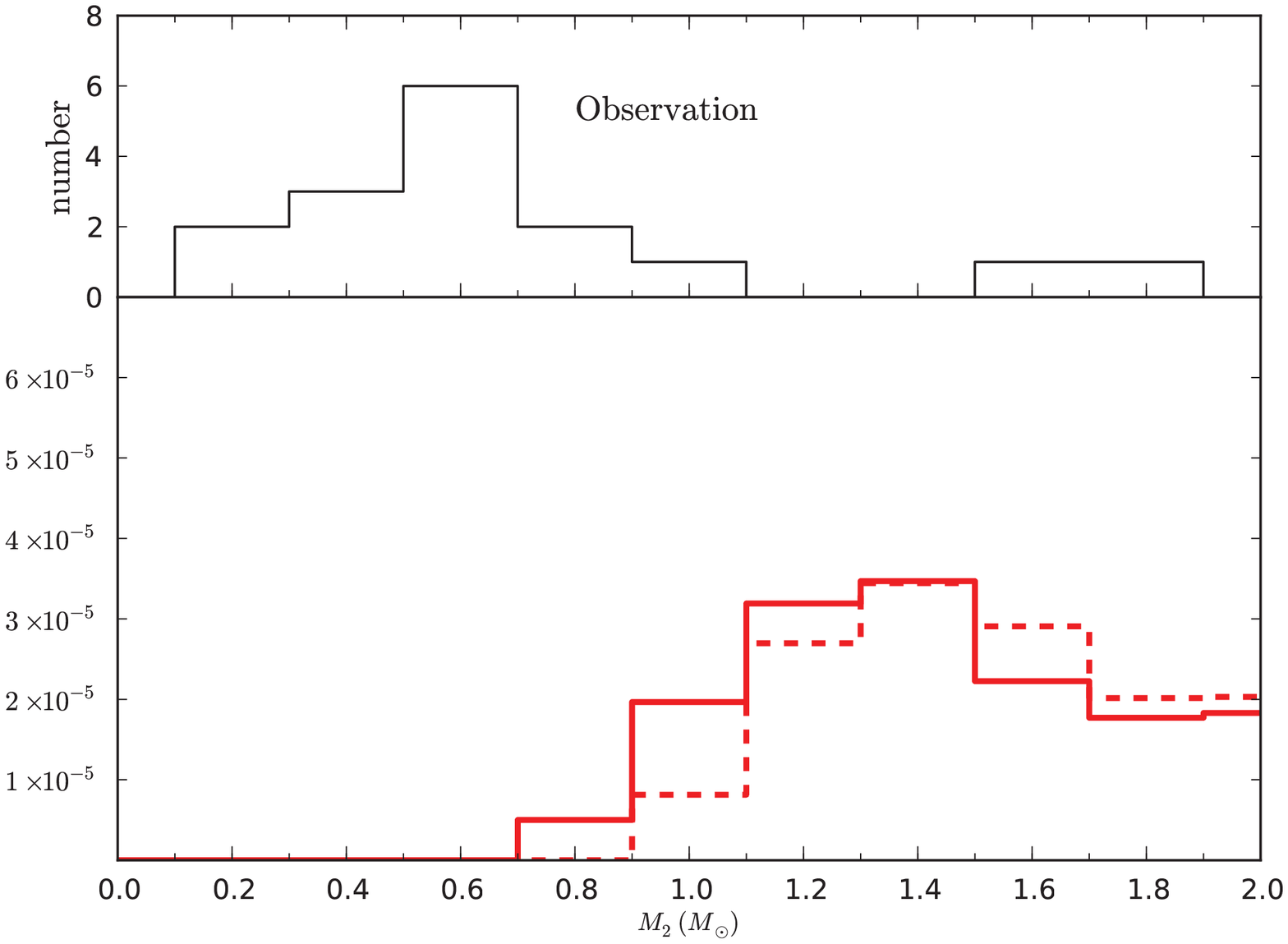}
          \end{minipage}

     \begin{minipage}[h,t]{3.3in}
     \includegraphics[width=3.3in]{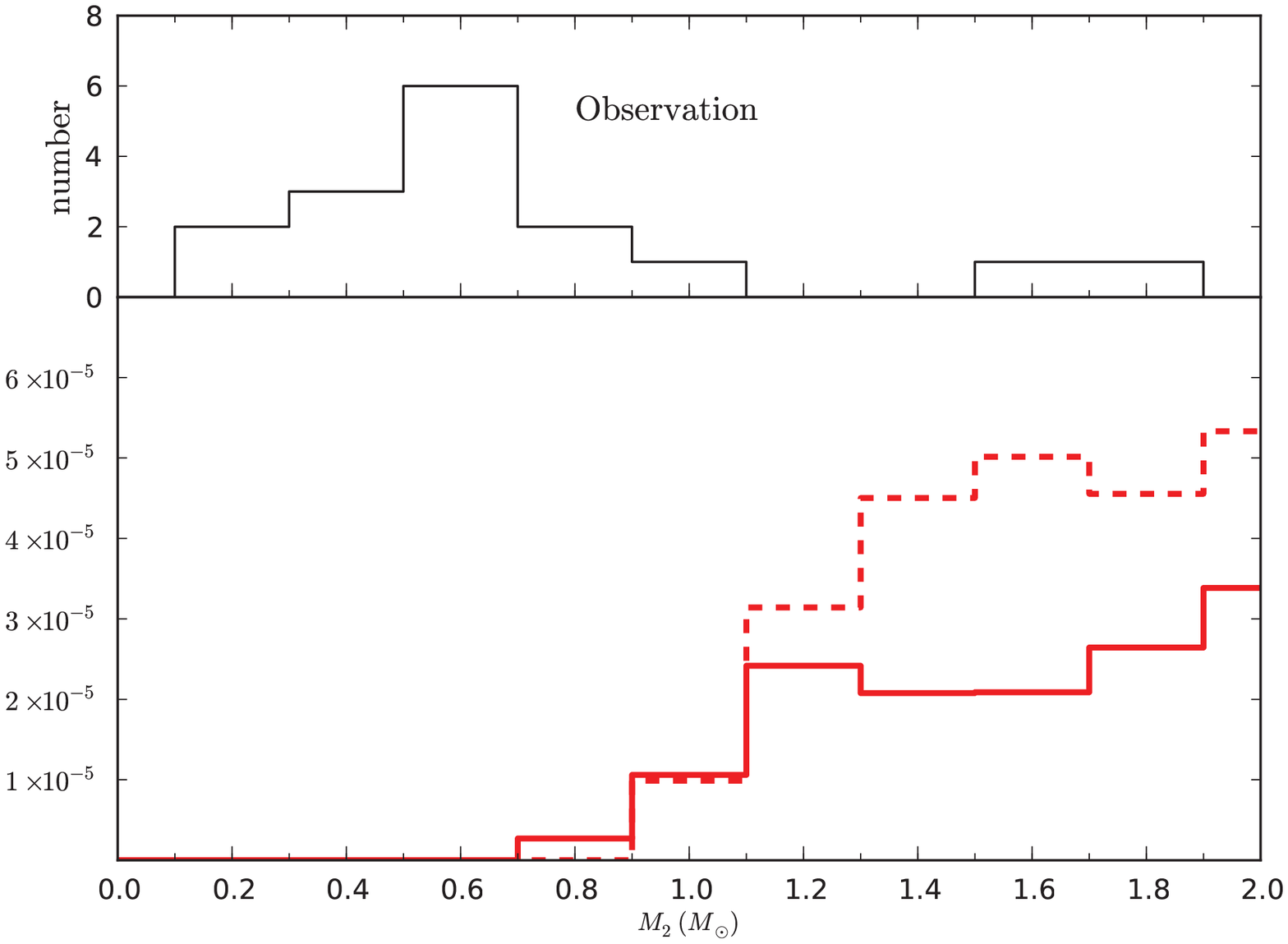}
     \end{minipage}
     \end{tabular}

\caption{Distributions of the donor mass for the BH XRBs in models
A1 (left) and A2 (right).  The solid and
dashed red lines stand for the results with the Wind1 and Wind2 prescriptions, respectively.
Also shown is the donor
mass distribution of observed LMXBs in the black line. \label{figure4}}
\end{figure}

We first consider the formation of BHs via the rapid SN mechanism.
In this case the progenitor stars are usually more massive than
$28\, M_{\sun}$ in binaries. For such massive stars the binding energy
of the envelope is so large that with $\lambda=\lambda_{\rm g}$  a
low-mass secondary inevitably merges with the He core of the
primary, and no successful BH LMXBs can form
\citep[see also][]{Podsiadlowski2003}. So in models A1 and A2 we set $\alpha_{\rm CE}=1$ with
$\lambda=\lambda_{\rm b}$ and $\lambda_{\rm h}$, respectively. The distributions of the donor mass of
the resultant BH XRBs at current epoch are plotted in the left and
right panels of Fig.~\ref{figure4}, respectively.
To reveal the donor mass distribution in different scale, the donor masses
are confined within $0-6\, M_{\odot}$ and
$0-2\, M_{\odot}$ in the upper and lower panels of the figure, respectively.
The solid and dashed red lines represent the results with the $\lambda$s obtained
with the Wind1 and Wind2 prescriptions, respectively. Also displayed is the
distribution of the observed BH LMXBs with the black line \citep[data taken
from][]{Wiktorowicz2013, Fragos2015}. The number of the BH
XRBs is calculated by multiplying the formation rate with the
duration in a specific stage.
We only display the sources which are subject to the thermal-viscous instability
in an X-ray irradiated accretion disk and ``assumed" to be transient
\citep{Lasota2001,Lasota2008}\footnote{We caution that, despite the thermal-viscous instability
seem to succeed in explaining the dichotomy between the transient and the
persistent  LMXBs averaged over a long
enough period of time \citep{Coriat2012}, it may not be the only mechanism, other mechanisms
(e.g., irregular mass injection from the low-mass, fully convective donor stars) can also be at work.}.
They include almost all BH LMXBs produced.
The exceptions are the systems comprised of intermediate-mass donors
or with extremely short periods, which are not relevant here.
In model A1 and A2, the donors of persistent sources locate in the mass range of
$4 - 6\, M_{\sun}$, and their total number is comparable with that of transient sources.
When $\lambda=\lambda_{\rm
b}$, most BH XRBs harbor an intermediate-mass donor star (with mass
$M>3 M_{\sun}$), implying that lower-mass secondaries are unlikely to
survive CE evolution. The corresponding formation rate of the BH XRBs
is too low ($\sim 2.80\times 10^{-8}$ yr$^{-1}$) to account for the
observed systems in the Galaxy. When $\lambda=\lambda_{\rm h}$, the
formation rate enhances a bit ($\sim 3.57\times 10^{-8}$ $\rm yr^{-1}$),
and the donor mass distribution peaks at around $4\, M_{\sun}$.
It is seen that in both cases the distributions significantly deviate from the observed one.
The calculated orbital period distributions are too long ($>$ 1 day)
to be consistent with observations (see Fig.~\ref{figure12} below). These results clearly
disfavors the rapid SN mechanism, so in the following we only
consider the failed SN mechanism.

\begin{figure}

     \begin{tabular}{cc}
     \begin{minipage}[h,t]{3.3in}
     \includegraphics[width=3.3in]{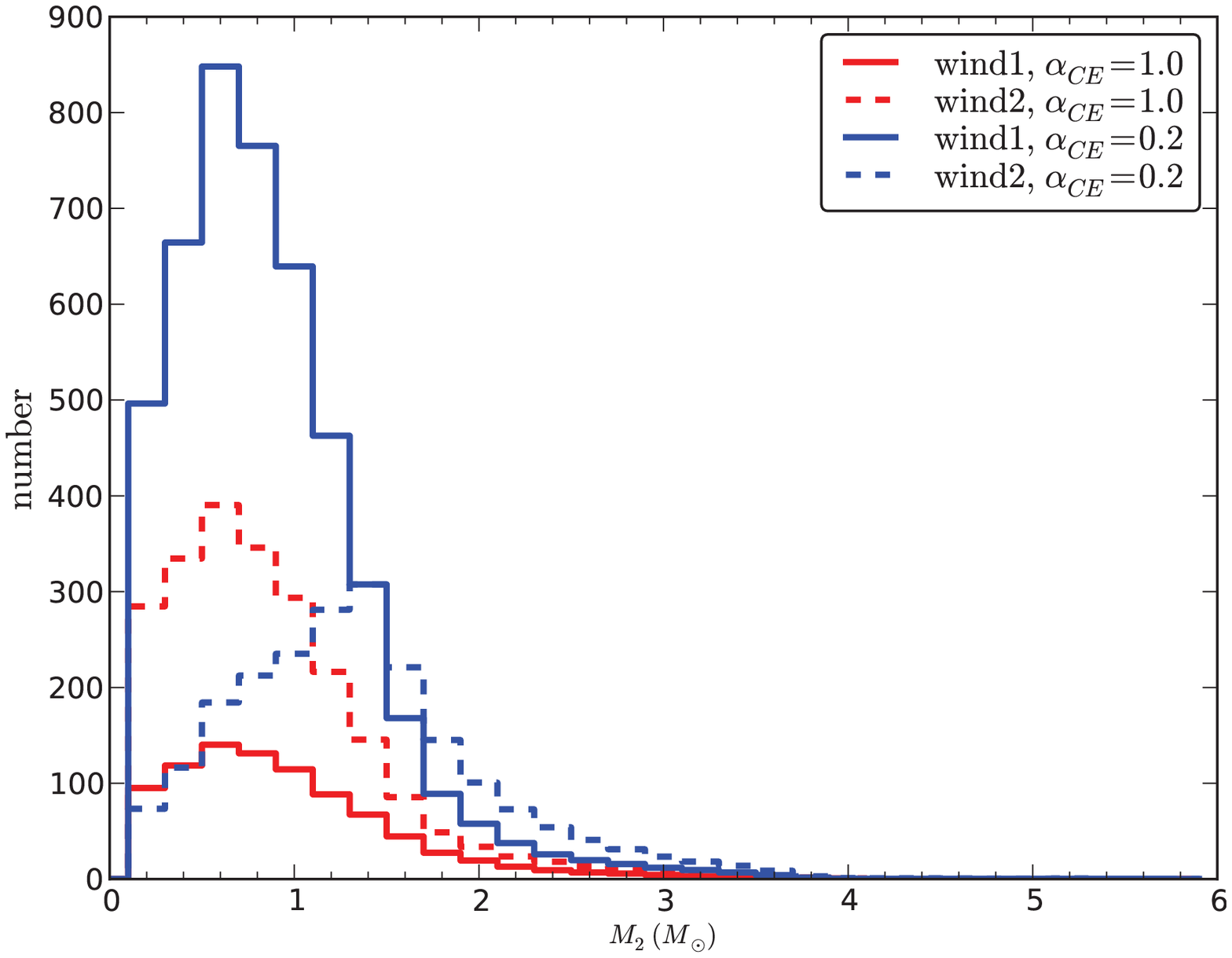}
     \end{minipage}
     \begin{minipage}[h,t]{3.3in}
     \includegraphics[width=3.3in]{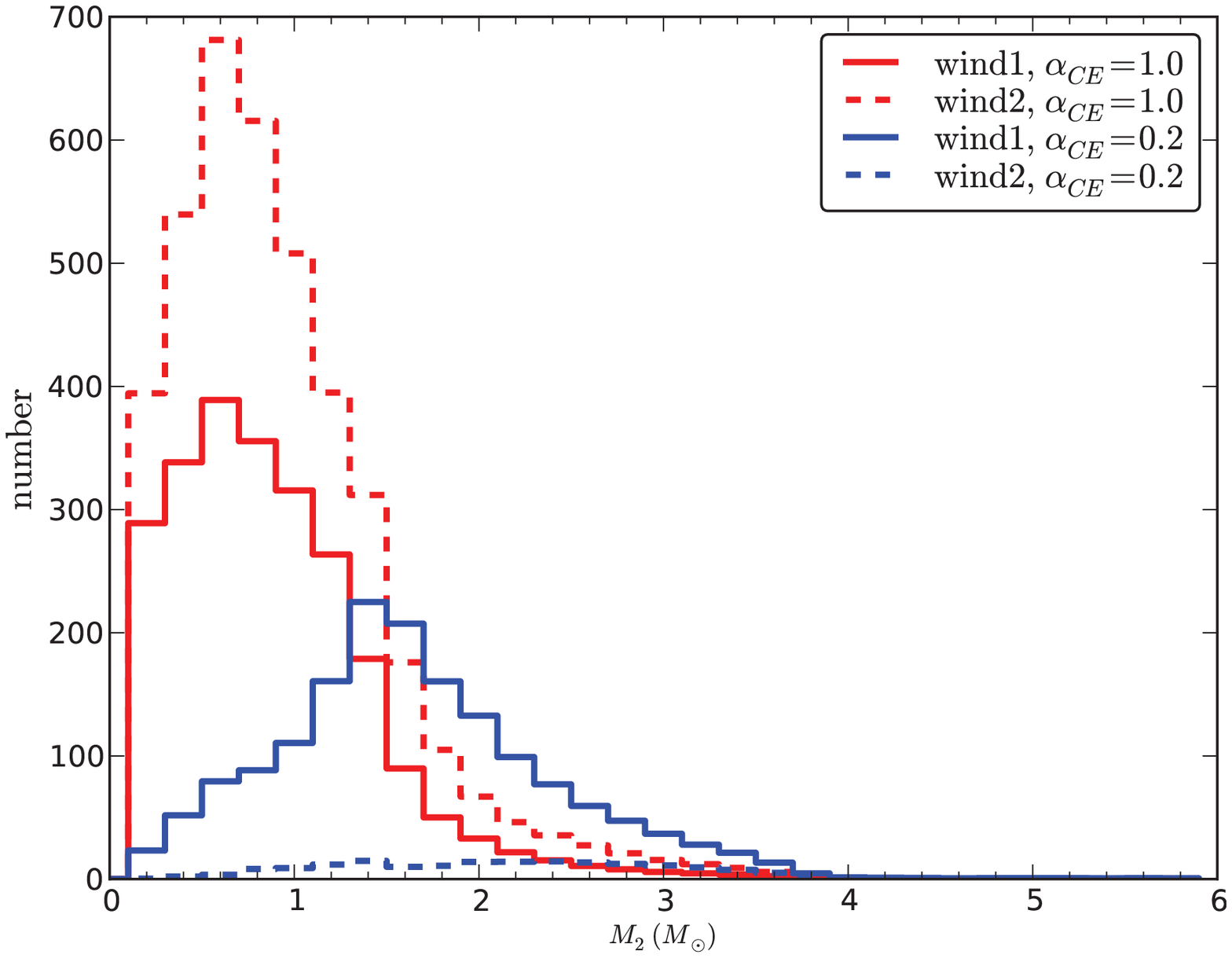}
     \end{minipage}\\
     \begin{minipage}[h,t]{3.3in}
     \includegraphics[width=3.3in]{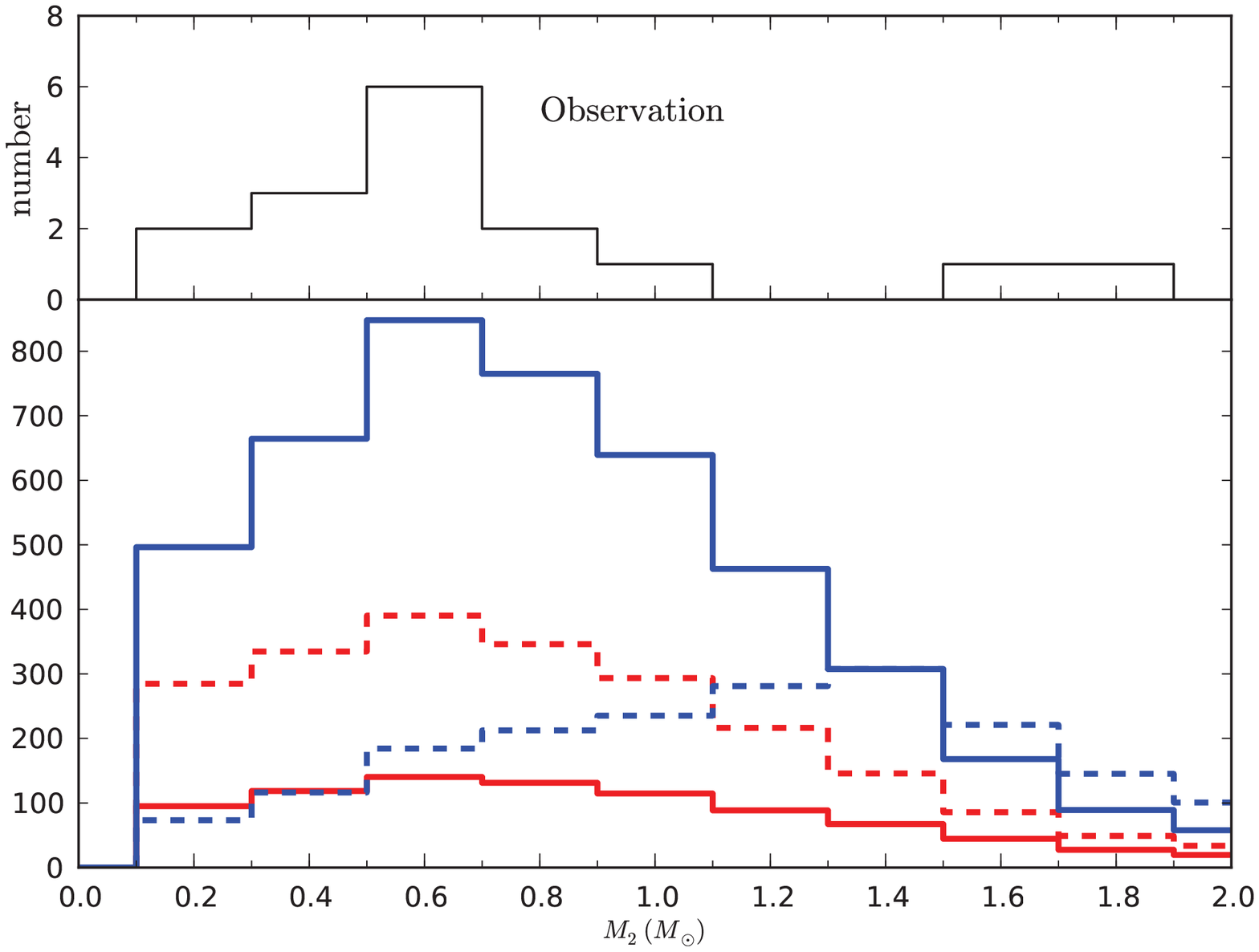}
     \end{minipage}
     \begin{minipage}[h,t]{3.3in}
     \includegraphics[width=3.3in]{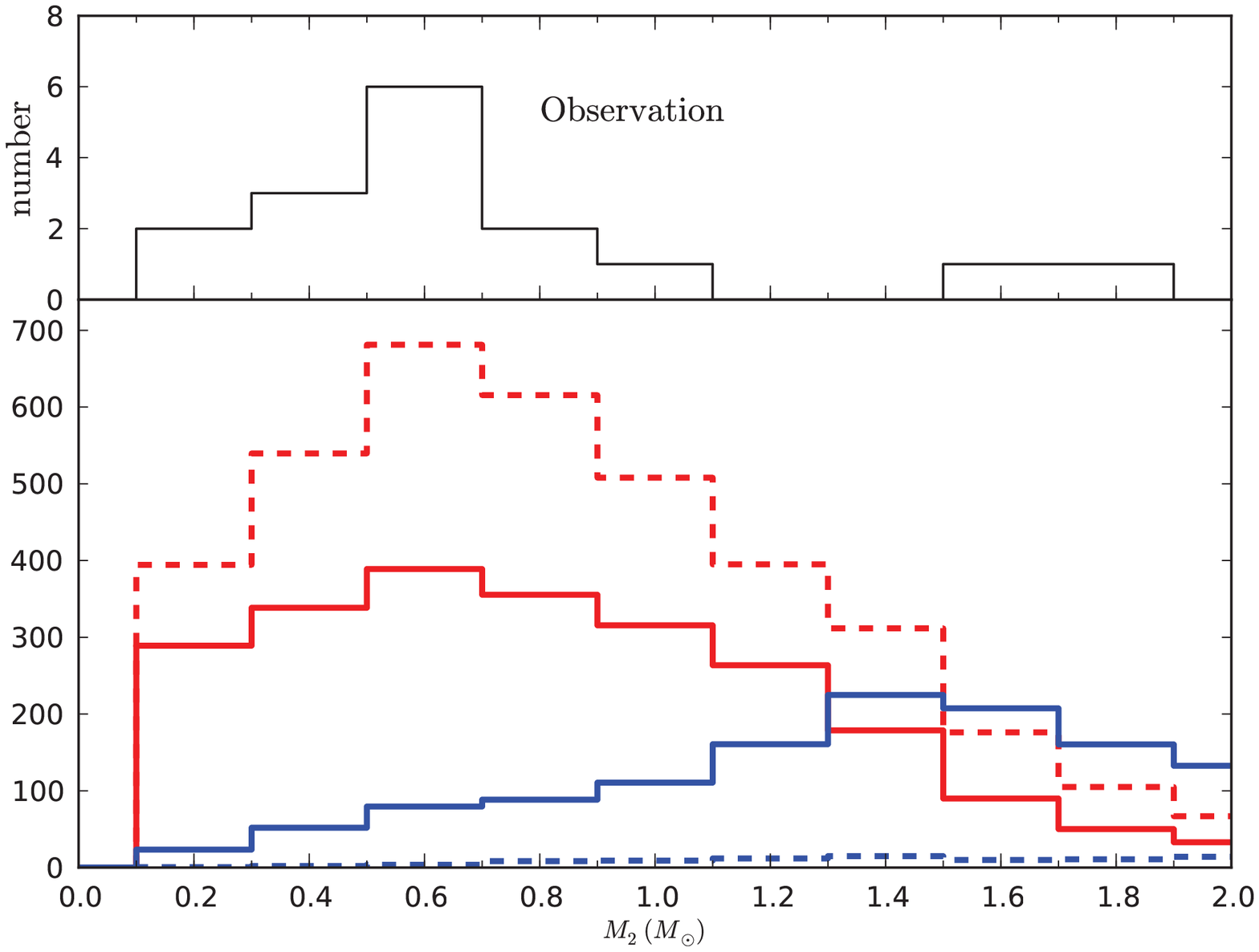}
     \end{minipage}
     \end{tabular}

\caption{Same as Fig.~3, but for models B1 (left) and B2 (right).
\label{figure5}}
\end{figure}

In models B1 and B2, we assume $M_{\rm BH}= \textit{M}_{\rm He}$ and a flat mass ratio distribution
with $\lambda =\lambda_{\rm b}$ and $\lambda =\lambda_{\rm g}$,
respectively. We present the calculated distributions of the donor mass
for the two models and compare them with observations in Fig.~\ref{figure5}
(left: B1; right: B2). Here the red and blue lines correspond to $\alpha_{\rm CE}=1$ and 0.2,
respectively, and the line styles are same as in Fig.~\ref{figure4}.
With the failed SN mechanism the masses of the BH progenitors can be
 lower than $20\, M_{\sun}$, so it is easier for low-mass secondaries to
survive the CE phase.
From the left panel of Fig.~\ref{figure5} we see that in the cases of Wind1/$\alpha_{\rm CE}=1$,
Wind2/$\alpha_{\rm CE}=1$, and Wind1/$\alpha_{\rm CE}=0.2$ the donor mass
peaks at $\sim 0.6\ M_{\sun}$, matching the observational data,
while in the case of Wind2/$\alpha_{\rm CE}=0.2$, the donor masses cluster around $1.2-1.4\ M_{\sun}$,
larger than the observed peak. This is simply because more massive donor stars are required
to survive the CE evolution when the product $\alpha_{\rm CE}\lambda$ is lower. However,
larger $\alpha_{\rm CE} \lambda$ does not always necessarily lead to
larger number  of BH LMXBs, since the formation of a BH XRB strongly depends on the orbital
separation just before and after the CE phase.
With the same initial orbital periods, larger $\alpha_{\rm CE} \lambda$ results in wider post-CE orbits,
which may cause the subsequent Roche lobe overflow to be postponed, reducing the X-ray lifetime and
thus the number of XRBs.

\begin{figure}
\centerline{\includegraphics[width=1.00\textwidth]{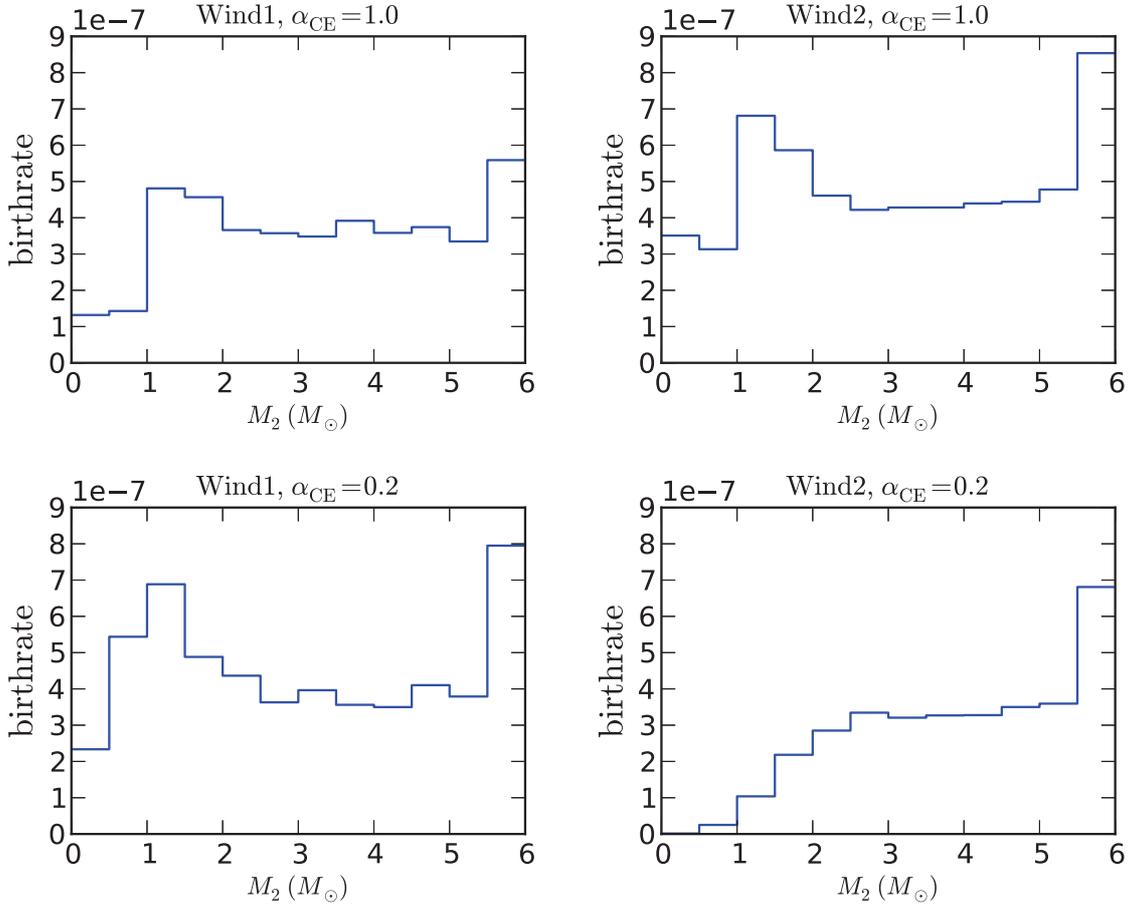}}
\caption{The birthrate of the BH XRBs as a function of the donor masses
for model B1 with different wind loss prescriptions and CE efficiencies.
\label{figure6}}
\end{figure}

Figure~\ref{figure6} shows the birthrates of the BH XRBs as a function of the initial donor mass
in each case.  We mention that, although the BH XRB birthrates can be as high as
$\sim 6\times10^{-6}$ yr$^{-1}$, the birthrates of those with incipient low-mass donors are several times lower.

We use the Kolmogorov-Smirnov (KS) test to
quantitatively  compare the the univariate distributions of the
calculated and measured donor mass. The KS statistic
$D$ quantifies a distance between the empirical distribution
functions of the two samples. The null hypothesis that the two
distributions are drawn from the same one is rejected at a
significance level $\alpha$ if $D>D_{\rm cr}(\alpha)$ \citep{fj12}.
In Table~\ref{table2} we list the calculated values of $D$ and $D_{\rm cr}(\alpha)$
for all models (except models A1 and A2 because
of too few data points) in the case of Wind2/$\alpha_{\rm CE}=1$ at
$\alpha=0.001$ and 0.05.

\begin{figure}

     \begin{tabular}{cc}
     \begin{minipage}[h,t]{3.3in}
     \includegraphics[width=3.3in]{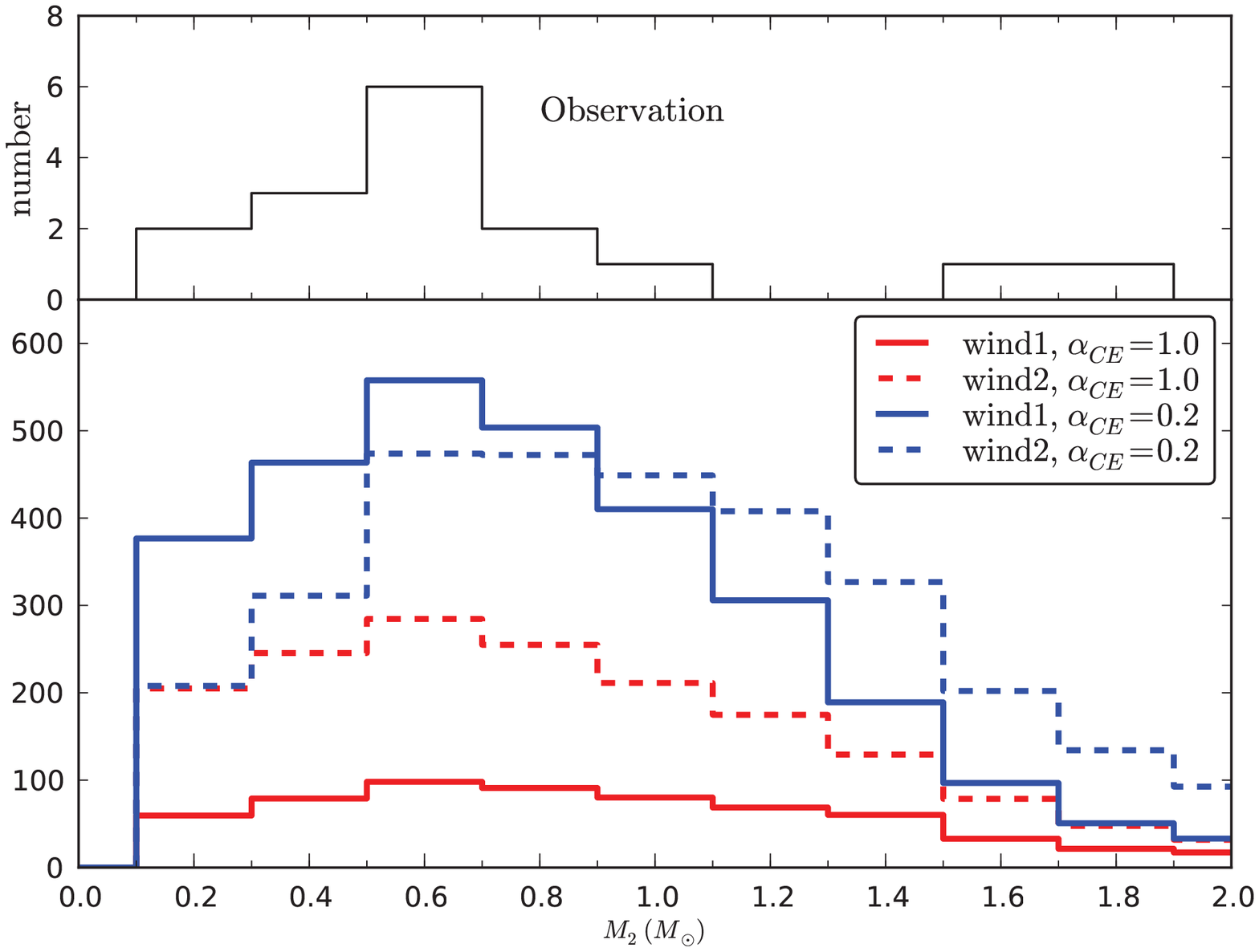}
     \end{minipage}
     \begin{minipage}[h,t]{3.3in}
     \includegraphics[width=3.3in]{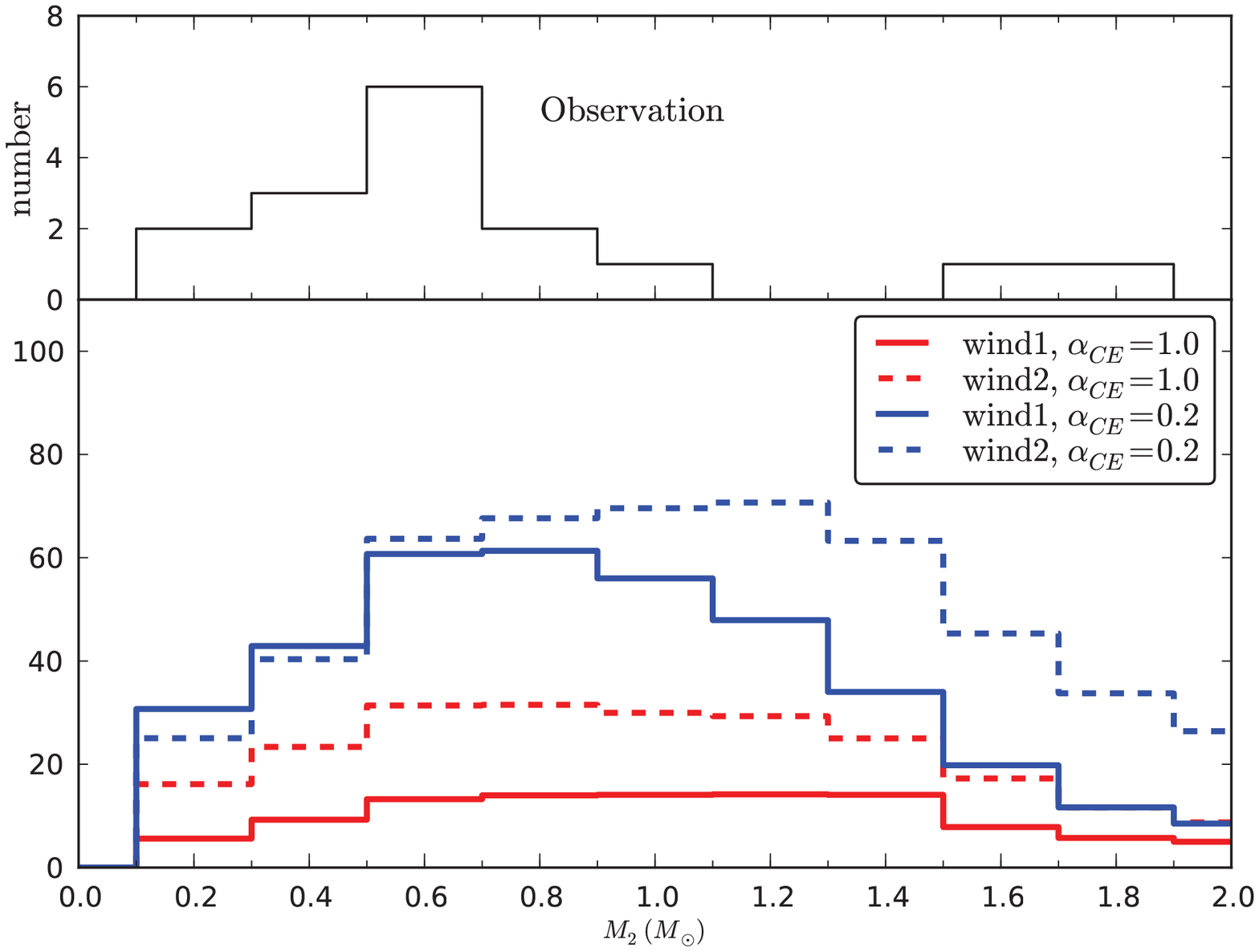}
     \end{minipage}\\

     \end{tabular}

\caption{Same as Fig.~3, but for models C1 (left) and C2 (right).
\label{figure7}}
\end{figure}

\begin{figure}
\centerline{\includegraphics[width=1.00\textwidth]{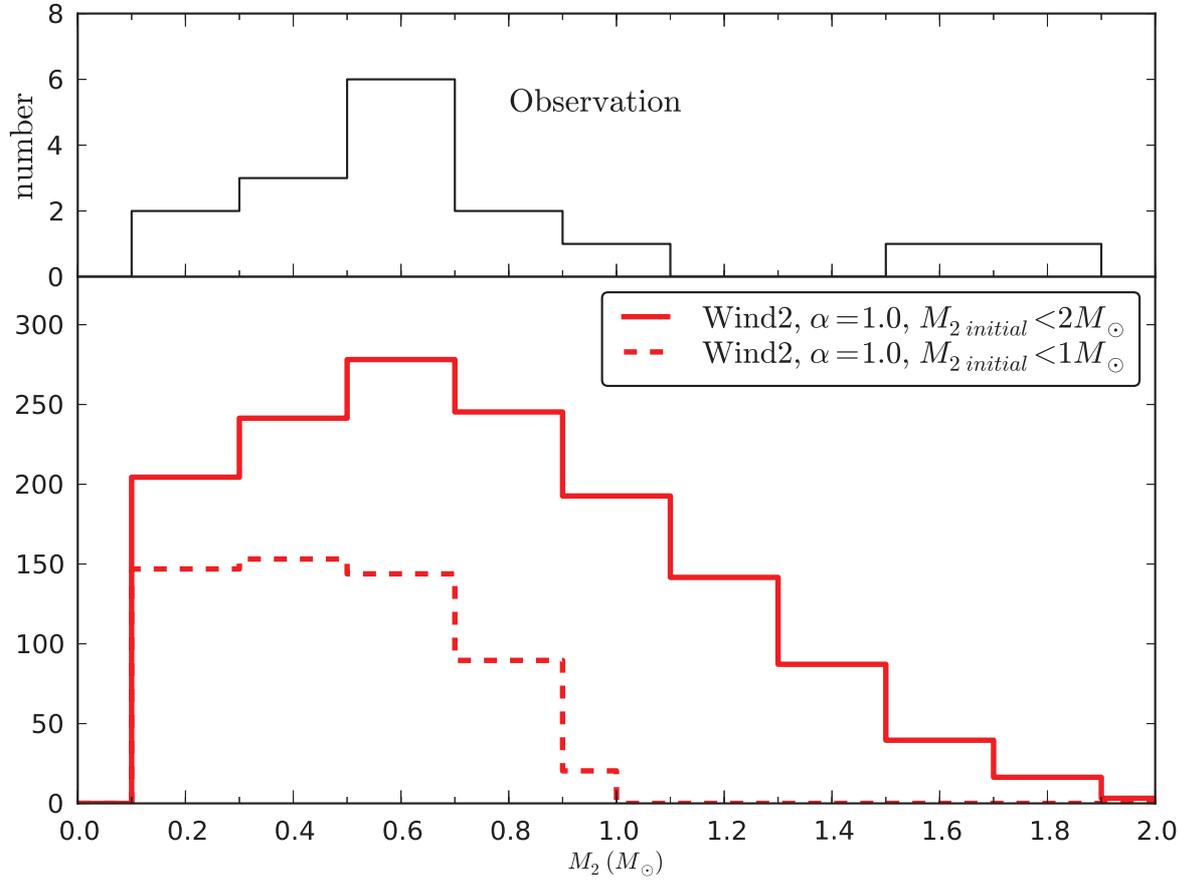}}
\caption{Distributions of the donor masses for BH XRBs in model C1.
The solid and dashed lines represent the binaries with initial donor mass
$<2\,M_\odot$ and $<1\,M_\odot$, respectively.
\label{figure8}}
\end{figure}

\begin{figure}

     \begin{tabular}{cc}
     \begin{minipage}[h,t]{3.3in}
     \includegraphics[width=3.3in]{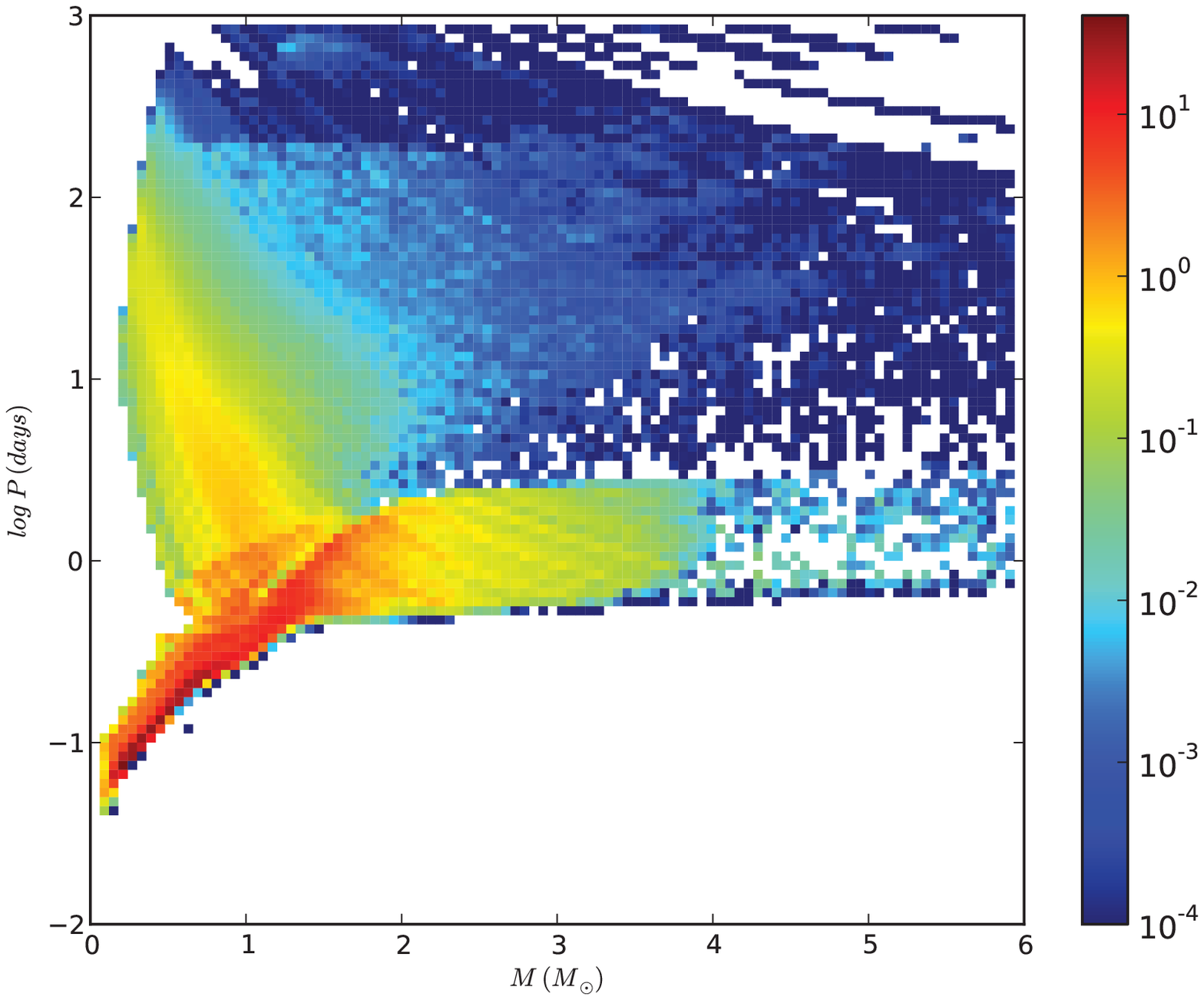}
     \end{minipage}
     \begin{minipage}[h,t]{3.3in}
     \includegraphics[width=3.3in]{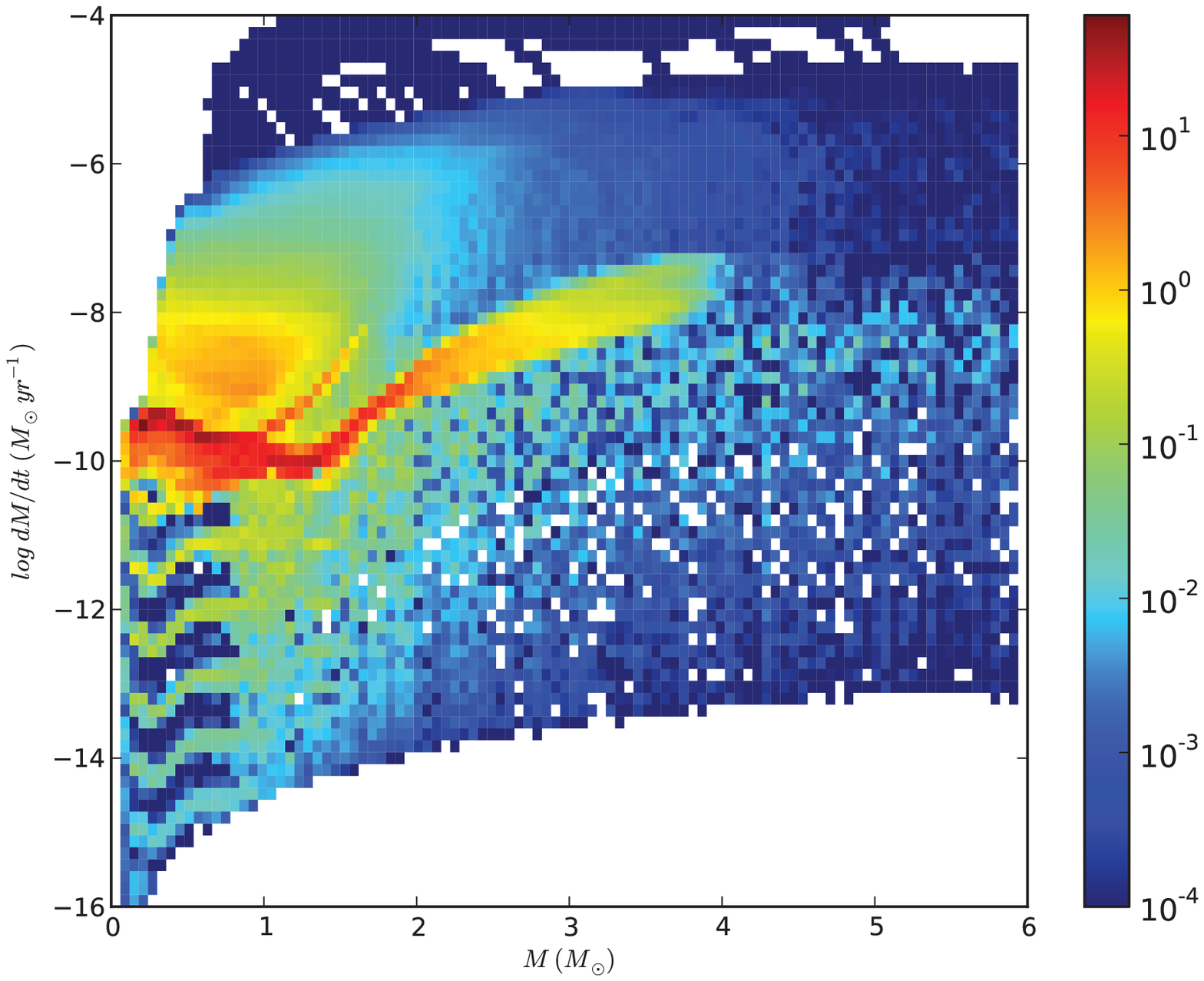}
     \end{minipage}\\

     \end{tabular}

\caption{Distributions of the orbital period vs. the donor mass (left panel) and the
mass transfer rate vs. the donor mass (right panel) in model C1.
The color scale indicates the relative number of XRBs.
\label{figure9}}
\end{figure}

Figure ~\ref{figure7} shows the donor mass distributions for models C1
(left panel) and C2 (right panel) under the assumptions that $M_{\rm
BH}=\textit{M}_{\rm He}$
 and $\lambda=\lambda_{\rm h}$. In models C1 and
C2 we adopt a flat distribution of the initial mass ratio and a ``twin"
distribution \citep[][see however, Cantrell \& Dougan
2014]{Pinsonneault2006,Kobulnicky2007}, respectively. Here we
confine the donor masses within the range of $0-2\, M_{\sun}$ because of the
small number of XRBs with more massive donor stars. Figure~\ref{figure7}
shows that the peak of the donor mass distribution in the cases of
Wind1/$\alpha_{\rm CE}=1$, Wind2/$\alpha_{\rm CE}=1$, and Wind1/$\alpha_{\rm
CE}=0.2$ is $\sim 0.6\,M_{\sun}$, but shifts towards $\sim 1.0\,M_{\sun}$ in
the case of Wind2/$\alpha_{\rm CE}=0.2$, roughly in accordance with
Fig.~\ref{figure5}. This is also reflected by their similar $D$ values. To
see whether the resultant BH LMXBs evolve from incipient low- or
intermediate-mass binaries, we repeat the calculation with the initial donor
mass confined to be $<2\ M_{\sun}$ and $<1\ M_{\sun}$ respectively. The
results in model C1 (with the Wind2 prescription) are plotted in
Fig.~\ref{figure8}. It is clearly seen that the total number of the BH LMXBs
evolved from the primordial binaries with the initial donor mass $M_{2
initial}\, < 2\, M_{\sun}$ is almost the same as in the left panel of
Fig.~\ref{figure7}, so we can safely conclude that almost all the BH LMXBs in
our calculation have incipient low-mass secondaries. Although the BH XRBs
with low-mass donors have a lower birthrate than those with
intermediate-mass donors, it is compensated by their long X-ray lifetimes.
Figure~\ref{figure7} also indicates that assuming a $n(q)\propto q$
distribution (model C2) leads to less efficient formation of BH LMXBs than
with a flat $q$ distribution because of less progenitor binaries with
low-mass secondary stars. Figure~\ref{figure9} shows the distribution of the
orbital period vs. the donor mass (left panel), and the distribution of the
mass transfer rate vs. the donor mass (right panel) in model C1. The color
scale represents the relative number of the XRBs in a parameter space. It
can be seen that the majority of BH XRBs are short-period ($<1$ day) systems
with  mass transfer rates below $10^{-9} M_{\sun}\rm yr^{-1}$.

\begin{figure}

     \begin{tabular}{cc}
     \begin{minipage}[h,t]{3.3in}
     \includegraphics[width=3.3in]{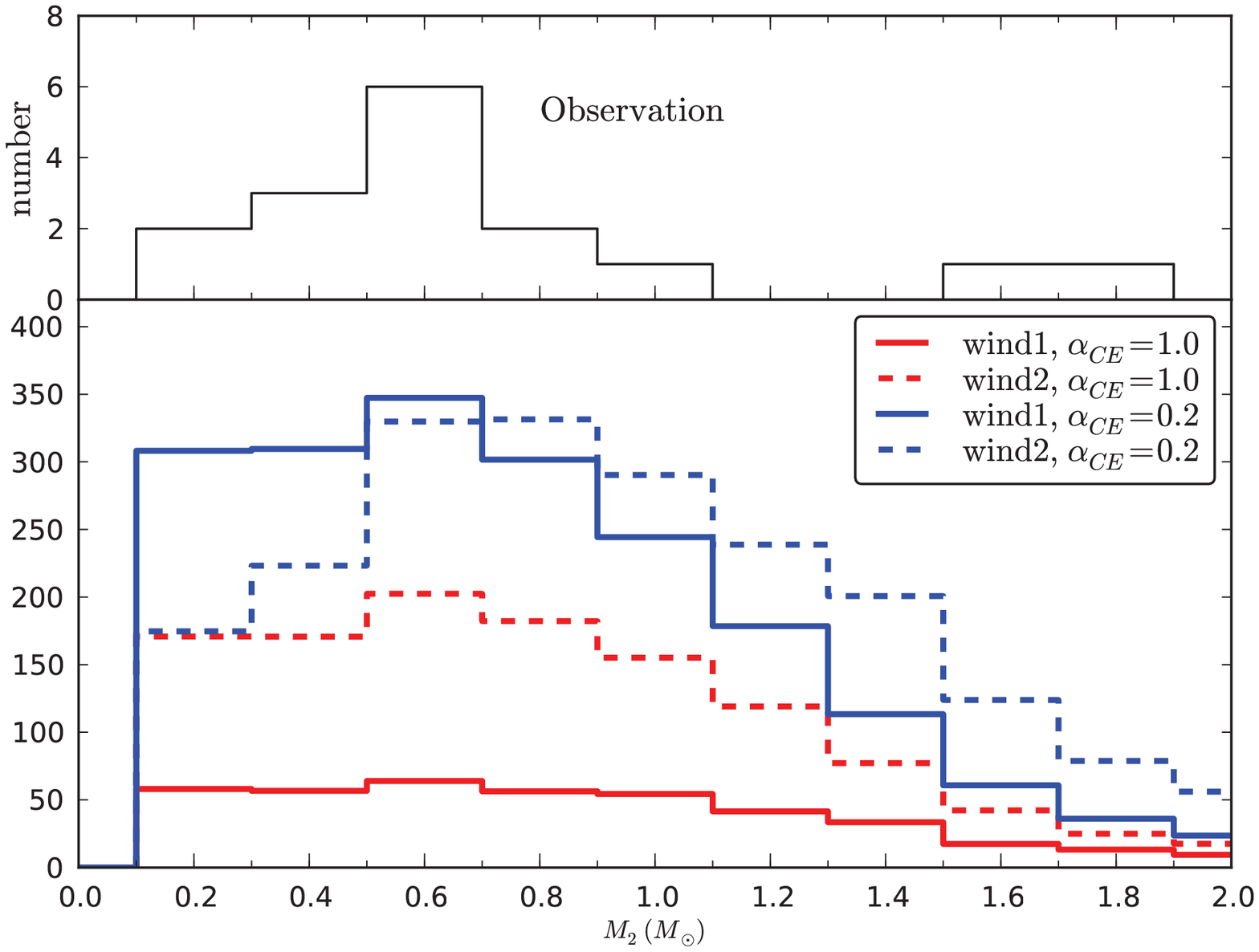}
     \end{minipage}
     \begin{minipage}[h,t]{3.3in}
     \includegraphics[width=3.3in]{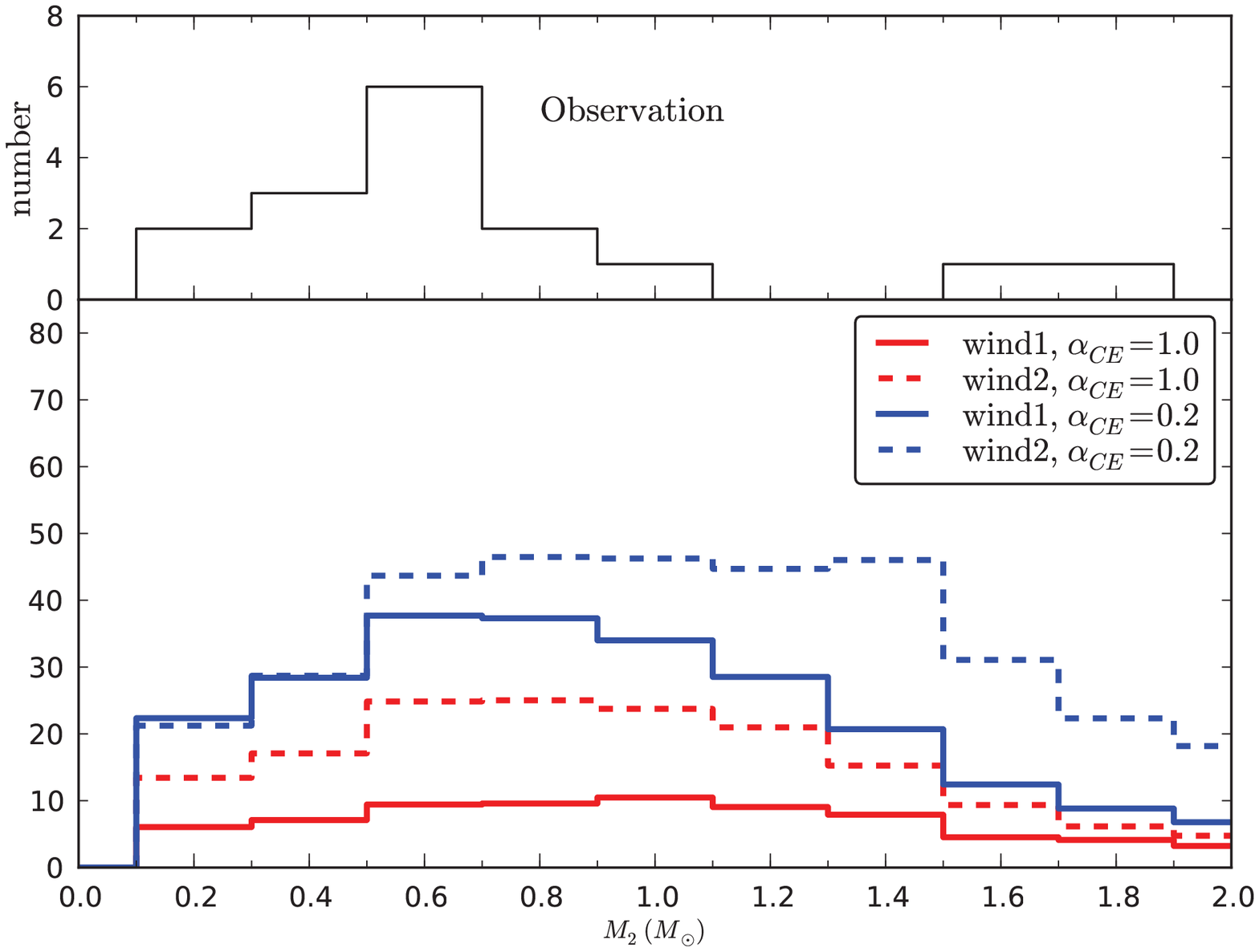}
     \end{minipage}\\

    \end{tabular}

\caption{Same as Fig.~3, but for models D1 (left) and D2 (right).
\label{figure10}}
\end{figure}

\begin{figure}
\centerline{\includegraphics[width=1.00\textwidth]{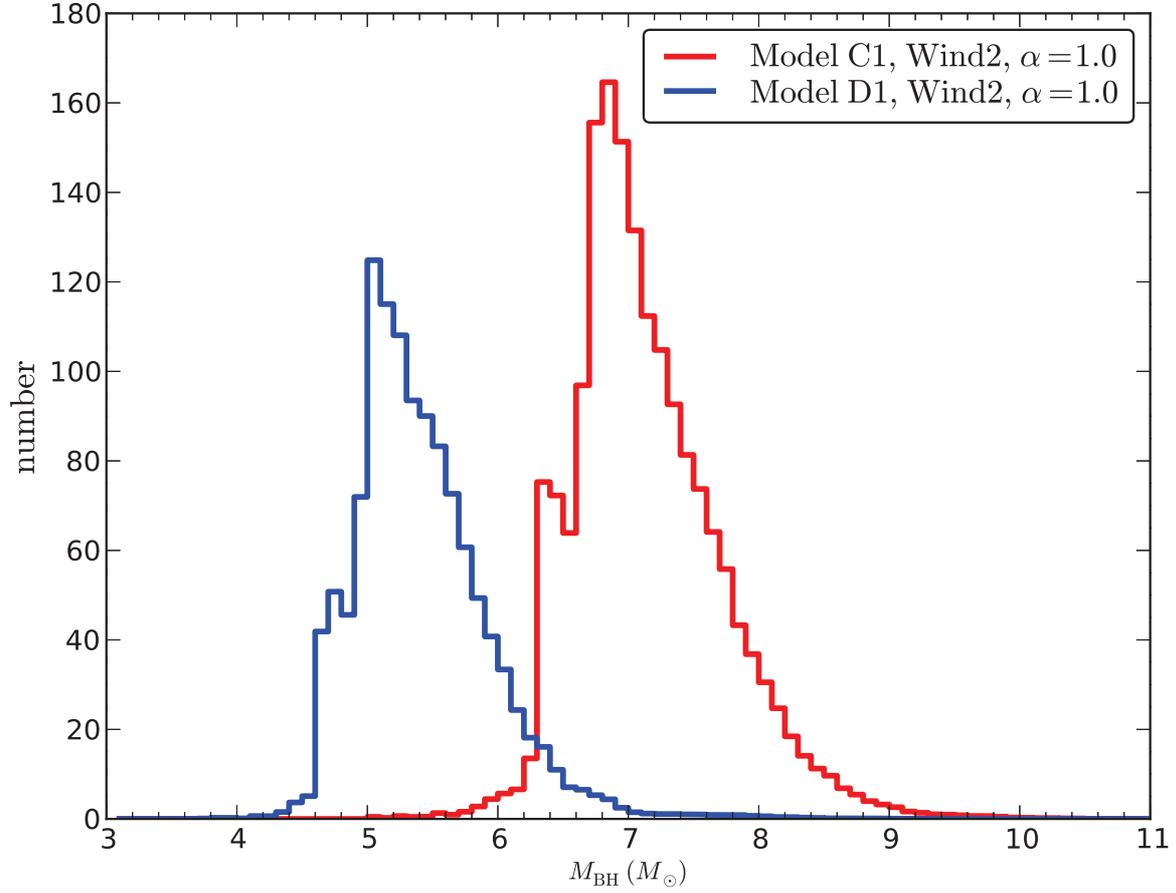}}
\caption{Distributions of the BH mass in models C1(red) and D1 (blue). \label{figure11}}
\end{figure}

The above results are based on the hypothesis that the BH mass is equal to
its progenitor's He core mass at the time of core bounce. To check the
effect of another choice of the remnant mass, we construct models D1 and D2,
which are same as models C1 and C2 except with $M_{\rm rem}=M_{\rm CO}$,
respectively. The donor mass distributions are shown in Fig.~\ref{figure10}.
We find that the formation rate in mode D1 decreases to about half
that in model C1, and the peak of the donor mass distribution extends to a
broader range compared with Fig.~\ref{figure6}. The remnant mass influences
the formation of the BH LMXBs by determining the Roche lobe size after the
SN explosion. Figure~\ref{figure11} shows the distributions of the BH masses
under the two choices. The red and blue lines represent the results in
models C1 and D1, which peak at $\sim 6.6-7\,M_\odot$ and $\sim
5.0-5.4\,M_\odot$, respectively. \cite{Ozel2010} have analyzed the data of
16 BH X-ray transients, and found that the observed BH masses can be best
described by a narrow distribution at $7.8\pm 1.2\, M_{\sun}$, which is
obviously more compatible with the result in  model C1.

\subsection{Distribution of the orbital period and the effective
temperature of the donor stars}

\begin{figure}

     \begin{tabular}{cc}
     \begin{minipage}[h,t]{3.3in}
     \includegraphics[width=3.3in]{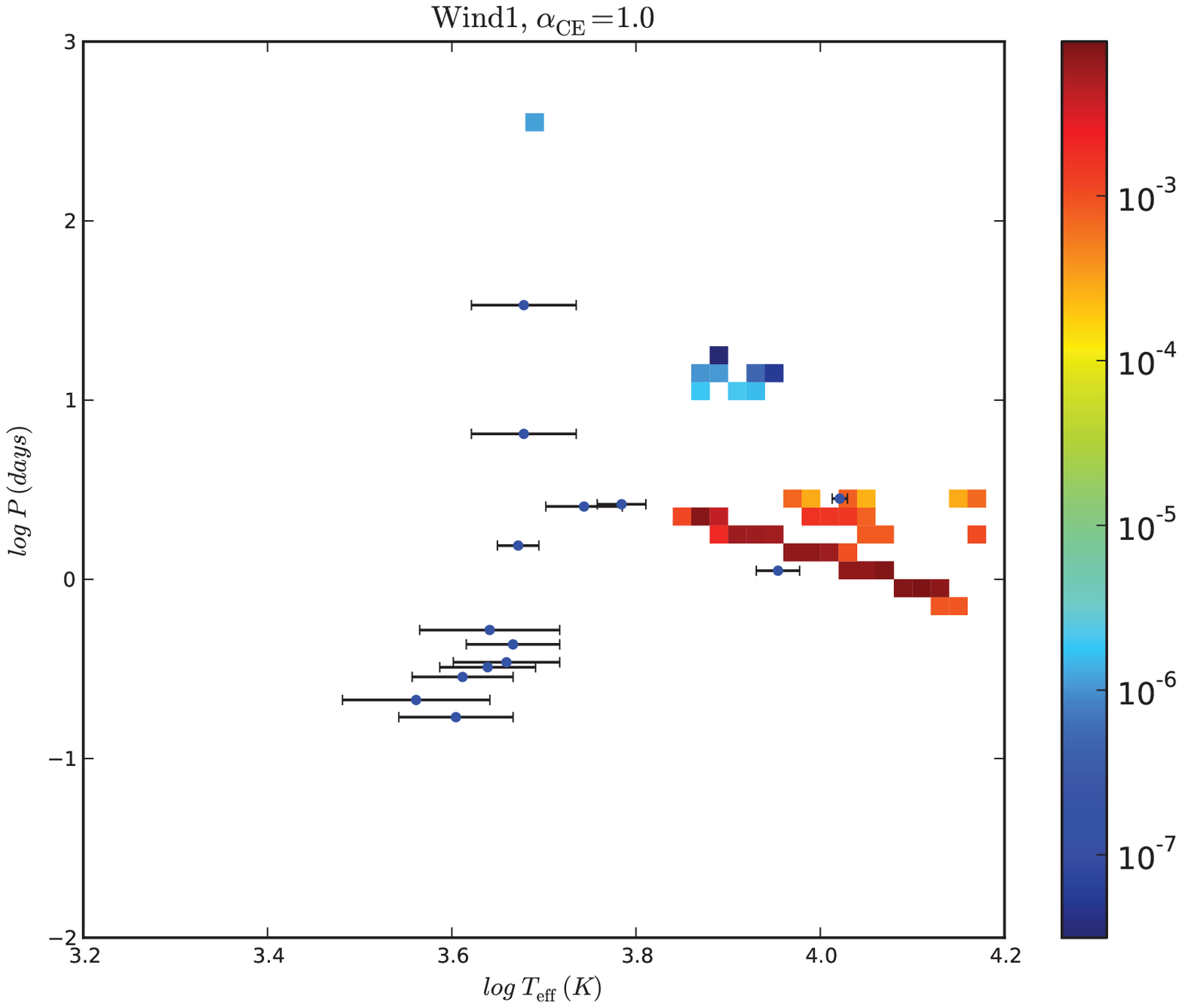}
     \end{minipage}
     \begin{minipage}[h,t]{3.3in}
     \includegraphics[width=3.3in]{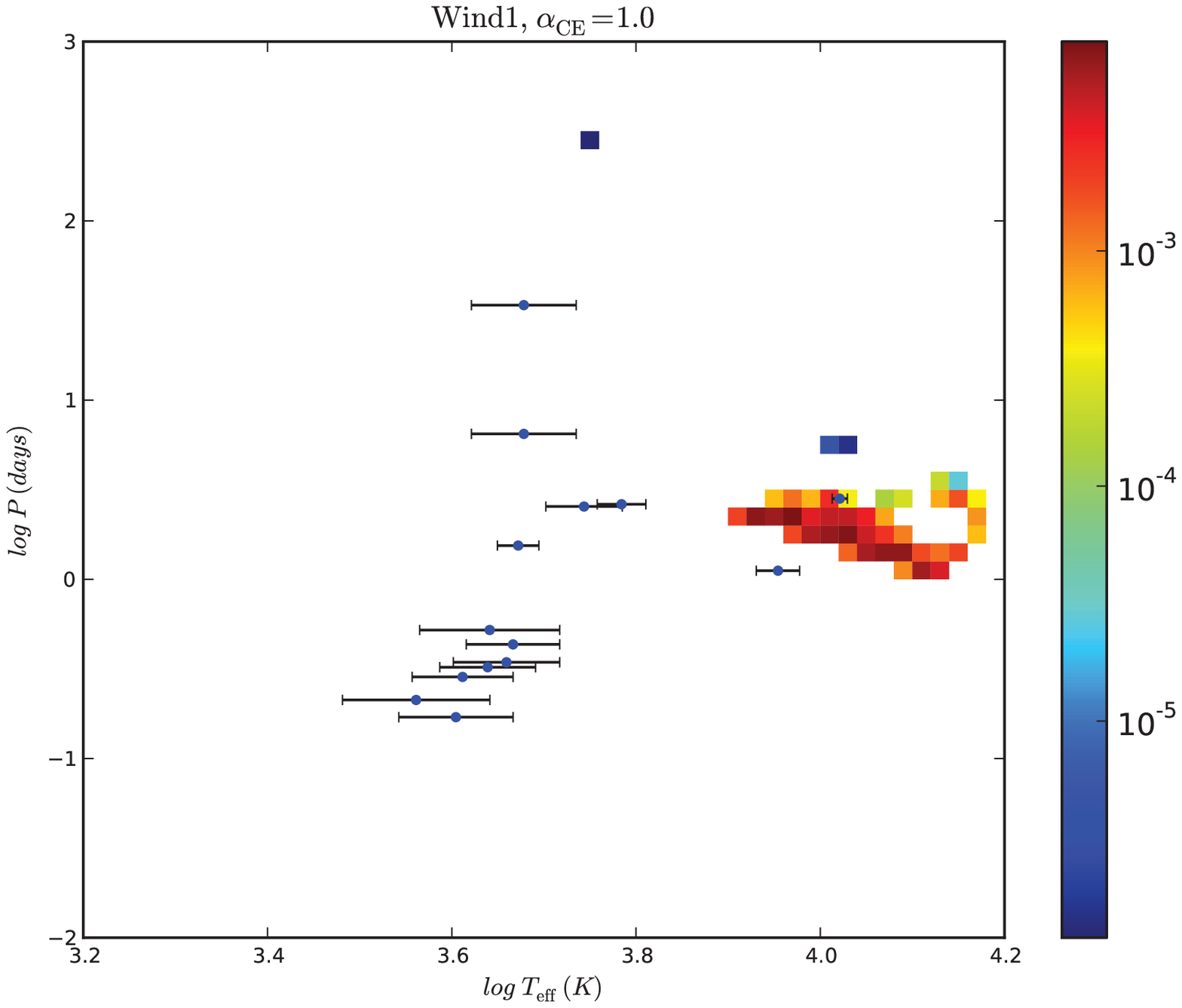}
     \end{minipage}\\
     \end{tabular}
\caption{Distribution of the orbital period and the donor star's
effective temperature for models A1 (left panel) and A2 (right panel)
with the Wind1 prescription and $\alpha_{\rm CE}=1$. The color scale reflects the
relative number of the binaries. For comparison, we also plot the
measured orbital periods and effective temperatures of Galactic BH
LMXBs.
\label{figure12}}
\end{figure}

\begin{figure}

     \begin{tabular}{cc}
     \begin{minipage}[h,t]{3.3in}
     \includegraphics[width=3.3in]{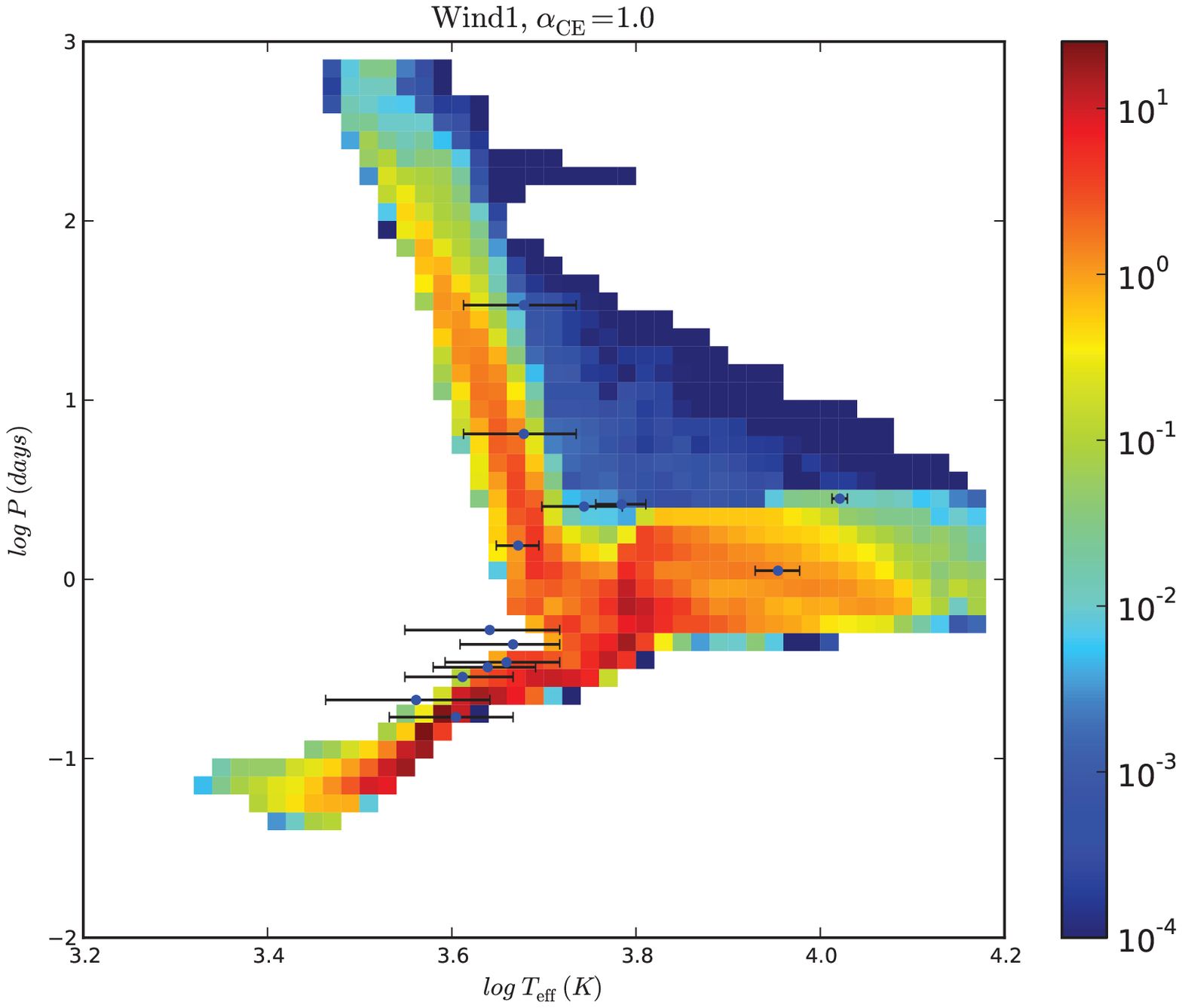}
     \end{minipage}
     \begin{minipage}[h,t]{3.3in}
     \includegraphics[width=3.3in]{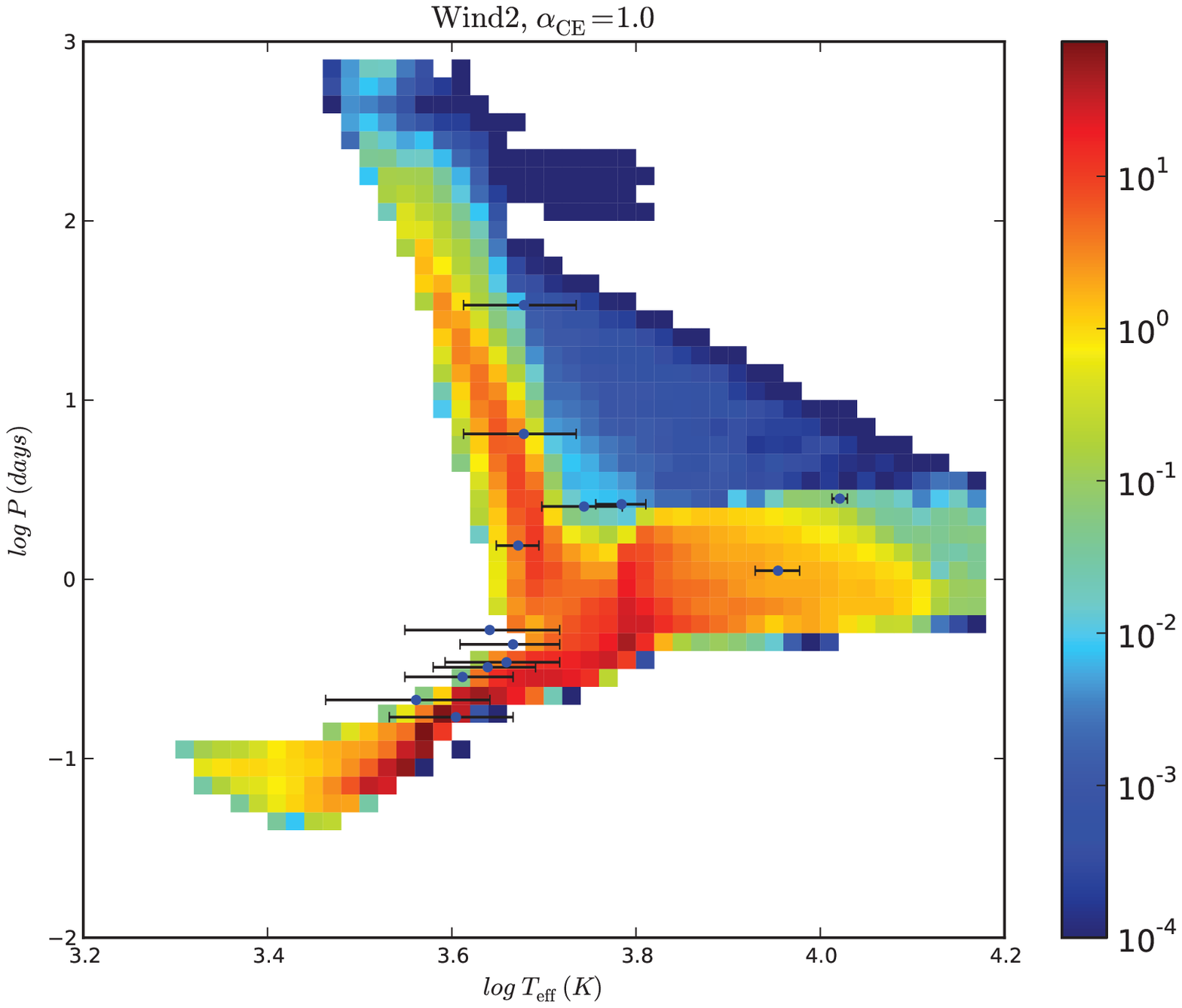}
     \end{minipage}\\
     \begin{minipage}[h,t]{3.3in}
     \includegraphics[width=3.3in]{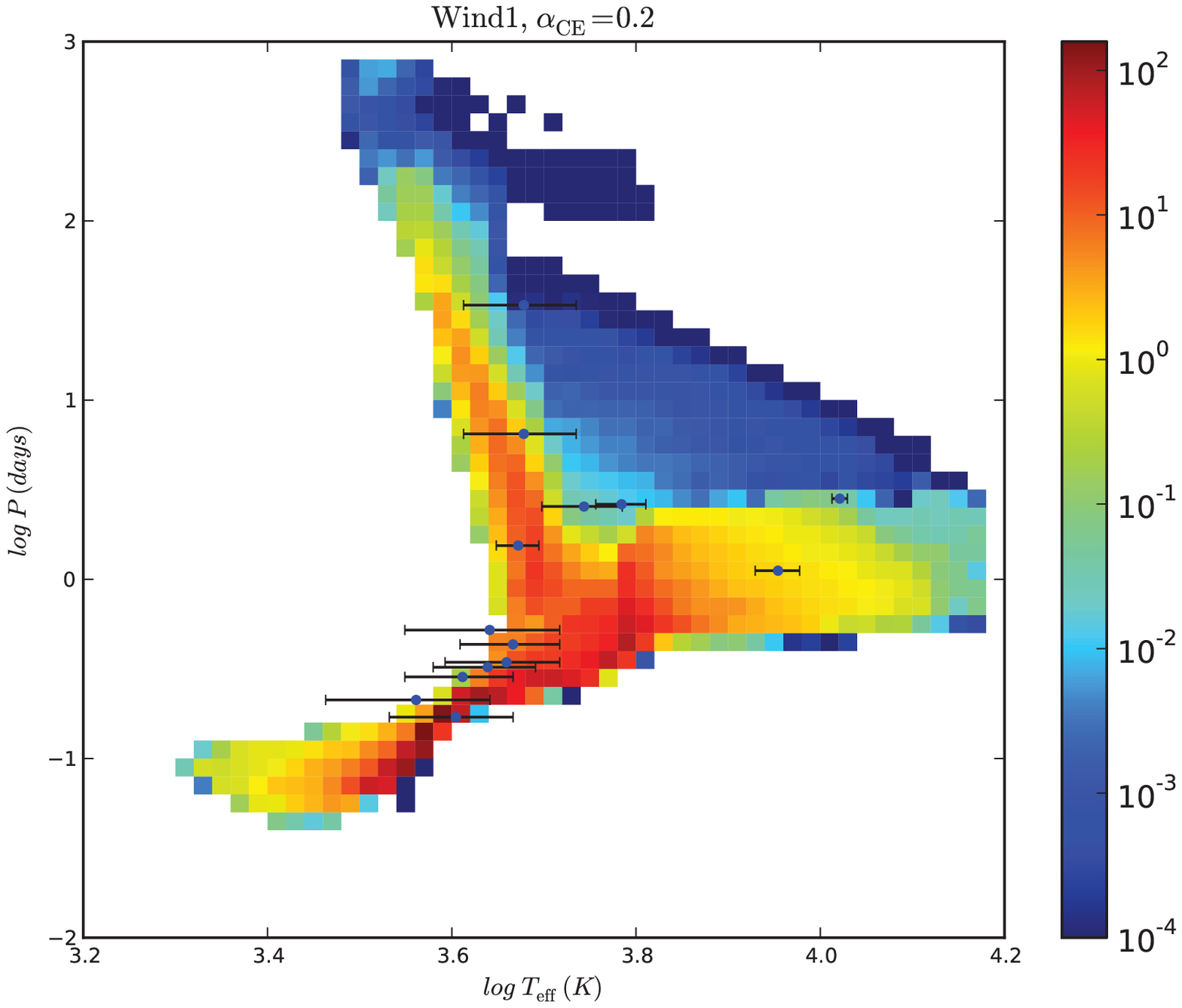}
     \end{minipage}
     \begin{minipage}[h,t]{3.3in}
     \includegraphics[width=3.3in]{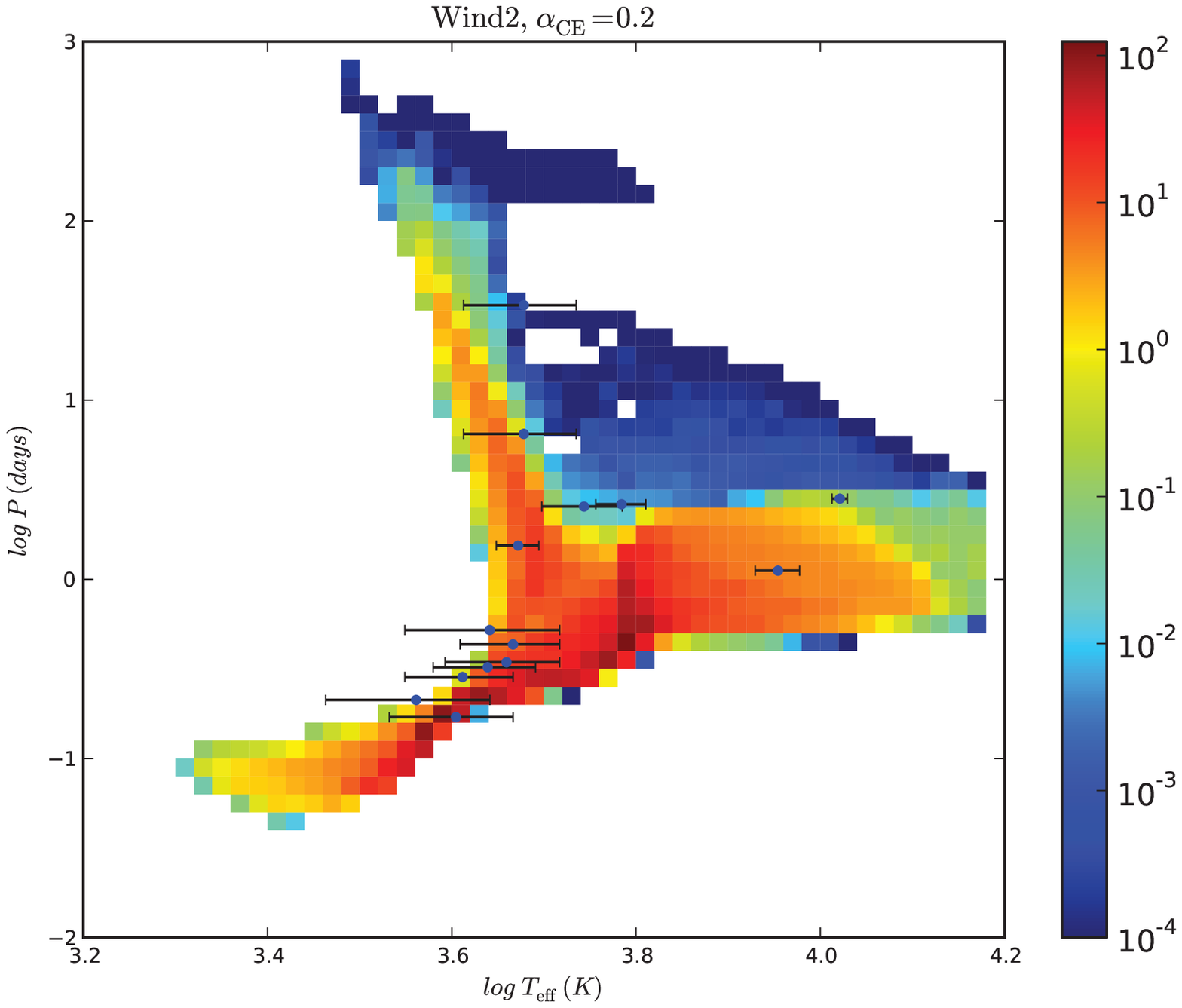}
     \end{minipage}
     \end{tabular}
\caption{Comparison between the measured and calculated orbital periods  and
donor star's effective temperatures in model C1 with different input parameters..
\label{figure13}}
\end{figure}

In Figs.~\ref{figure12}  we show the calculated distributions
of the orbital period and the effective temperature of the donor stars in
model A. The color scale represents the relative number of the XRBs.
We also plot the measured orbital periods and
effective temperatures of the donors for Galactic BH LMXBs. In
Fig.~\ref{figure12} for models A1 (left panel) and A2 (right panel), the observed distribution
obviously deviates from the calculation, showing a tendency towards lower
temperature. The reason is that, as stated before, most of the BH LMXBs in
models A1 and A2 descended from IMXBs, so the donor star is likely to have a
higher temperature than an originally low-mass star. We get similar results
for model B2 with $\alpha=0.2$. In other models the calculated
orbital periods are more compatible with observations.
We also use the KS test to quantitatively compare the calculated
and observed distributions, and the results for models B1, B2, C1, and D1 are
shown in Table~\ref{table3}. We can see that, except model B2 with
$\alpha_{\rm CE}=0.2$, other models are all acceptable.
We show the distributions in the best-fit model C1
in Fig.~\ref{figure13}.

\section{Discussion and conclusions }

Previous studies have already demonstrated that it is hard to
produce compact BHXRBs consisting of an initially low-mass companion
with the standard formation scenario, since the companion star is
unable to expel the CE \cite[e.g.,][see however, Yungelson \& Lasota
2008; Wiktorowicz et al. 2013]{por97,Kalogera1999,Podsiadlowski2003,KielHurley2006}. However,
the orbital periods, the donors' spectral types, and the measured
spins of quite a few BHs strongly suggest that in the incipient
binaries the donor stars are likely to be of low-mass.

We construct a series of models to explore the plausible solutions
to the above-mentioned puzzle. The factors we have taken into
account include the binding energy parameter $\lambda$, the CE
efficiency $\alpha_{\rm CE}$, the SN mechanisms for BH formation,
and the distribution of the mass ratio. By adopting different
choices of these prescriptions, we examine their roles in the
formation of BH LMXBs by comparing the calculated parameter
distributions with observations. Our calculations show that the
$\lambda$-values strongly depend on the stellar mass and the
evolutionary stage. Generally $\lambda_{\rm h}$ is several times larger than
$\lambda_{\rm g}$ and $\lambda_{\rm b}$, and this can help the
binary to avoid merging during the CE evolution without invoking
unphysically high values of $\alpha_{\rm CE}$. However, we fail to
reproduce the observational characteristics of Galactic BH LMXBs
based on the rapid SN mechanism even with $\lambda_{\rm h}$ adopted,
since the BH progenitors are still too massive ($\gtrsim
28\,M_{\sun}$) in this scenario. We then consider the alternative failed SN
mechanism, and assume that the remnant BH mass equals the He or CO
core mass of the primary. This time we
obtain more satisfactory results, with the distributions of the donor mass,
temperature and orbital period for the current binaries largely consistent with observations.
In particular, most BH LMXBs in this case originate from primordial binaries with
incipient low-mass ($<2\ M_{\sun}$) secondary stars,
compatible with the requirement derived from the spin
estimates of several BHs in LMXBs. Finally, all models predict
that most BH LMXBs are transient sources due to thermal
instability in the accretion disks. Thus, in the framework of the failed SN mechanism,
it is possible to reproduce the Galactic BH LMXBs with normal value of
$\alpha_{\rm CE}= 1$.

Realizing the uncertainties in the treatment of CE evolution, we also check
the results by adopting a small  $\alpha_{\rm CE}=0.2$ in each case. Our results
demonstrate that, when $\alpha_{\rm CE} \lambda$ is large enough for a low-mass
star to survive the CE phase, the donor mass distribution is not sensitive to the choice
of $\lambda$ or $\alpha_{\rm CE}$, which affects the number and birthrate of the produced
BH LMXBs.
However, for very small $\alpha_{\rm CE} \lambda$, the donor mass distribution strongly depends
on the value of $\lambda$ or $\alpha_{\rm CE}$, because surviving the CE phase becomes
paramount and difficult. In addition, the prescription of the BH mass with the He core mass
leads to better match of the observations than with the CO core mass.

Our results provide useful constraints on the formation process of
BH LMXBs, but are subject to many uncertainties in both
observational and theoretical aspects. When comparing the calculated
results with observations, one needs to be cautious that this
comparison is influenced by small number statistics as well as
selection effects. Currently there are around 20 dynamically
confirmed stellar-mass BHs. It is not known whether they are
correctly representative of the overall population in the Galaxy.
Most BH LMXBs are transient sources, and we are lack of thorough
understanding of their outburst characteristics (including the peak
luminosity and its relation with other parameters, and the duty
cycles) and the physical mechanisms \citep{RemillardMcClintock2006}.
To make things more complicated, a new class of very faint X-ray
transients were recently discovered with peak luminosities of only
$\sim 10^{34}-10^{36}$ ergs$^{-1}$ \citep{Wijnands2006}. Further
more, considering the selection effect that luminous sources are
more likely to be observed, we may expect that a large fraction of
the LMXBs possess even lower-mass donor stars than the current
sample, because they are relatively dim in X-rays even during
outbursts.

There are also big uncertainties in modeling the evolutionary
sequences of compact star binaries. We take into account $\lambda_{\rm
h}$ in calculating the binding energy of the envelope \citep[see
also][]{Wong2014}, but its credibility is controversial and there
are arguments both for and against it in the literature, let alone
the definition of the core-envelope boundary from which the binding
energy is integrated \citep[see][for a discussion]{Ivanova2013}. It
is also highly uncertain whether the CE efficiency $\alpha_{\rm CE}$
is constant or depends on other physical parameters.
\citet{DeMarco2011} and \citet{Davis2012} suggested that it may
decrease with increasing mass ratio, but this effect has not been
confirmed in modeling the SDSS post-CE binary population
\citep[e.g.,][]{Zorotovic2011} and needs further verification. Even
if true, the physics behind it is not clear, and numerical simulations in 3D
of CE evolution are eagerly required to resolve this issue. As to the
SN and the BH formation mechanisms, our results base on the
suggestion that stars of mass $\sim 17-25M_{\sun}$ die in failed SNe
leaving a BH \citep{Kochanek2014,Horiuchi2014}. The model still
needs to be tested by both 3D numerical simulations of the collapse
of massive stars and targeted field surveys for SNe in nearby
galaxies.

Obviously a thorough investigation on the evolutionary history of BH
LMXBs needs to incorporate BPS with the aforementioned input physics
in a self-consistent way.

\section*{Acknowledgements}

We are grateful to an anonymous referee for helpful comments.
This work was supported by the Natural Science Foundation of China
under grant numbers 11133001 and 11333004, and the Strategic
Priority Research Program of CAS (under grant number XDB09000000).







\newpage

\clearpage

\clearpage

\clearpage

\clearpage

\clearpage

\clearpage

\clearpage

%

\clearpage
\clearpage

\begin{table}

  \caption{The stellar parameters for a $20\,M_{\odot}$ star at different evolutionary stages.
Columns 2-4 and 5-7 correspond to the Wind1 (left) and Wind2 (right) prescriptions,
respectively \label{table1}}
  \begin{tabular}{ccccccc}
  \hline
  $ R\, (R_{\sun})$  & $ M\, (M_{\sun})$  &   $M_{\rm core}\, (M_{\sun})$   & $\lambda_{\rm h}$ & $ M\, (M_{\sun})$   &   $M_{\rm core}\, (M_{\sun})$   & $\lambda_{\rm h}$ \\
  \hline
  52&19.414&5.504&0.675&18.830&5.535&0.602\\
  600&18.738&5.984&0.075&18.787&5.580&0.077\\
  1066&14.098&6.548&0.152&11.72&6.801&0.027\\
  1230&14.096&6.516&0.389&11.54&6.760&0.126\\
  \hline
  \end{tabular}

\end{table}

\begin{table}
\caption{Input and derived parameters in the adopted models}\label{table2}
\begin{tabular}{lcccccccc}
\hline Models & $M_{\rm rem}$  & $\lambda$  & $n(q)$    &  birthrate (yr$^{-1}$) & $D_{\rm cr}(0.001)$ &
$D_{\rm cr}(0.05)$
& $D$ \\
\hline
A1 & Eq.~(8) & $\lambda_{\rm b}$ &    1   &  $2.80\times 10^{-8}$  &  &  \\
A2 & Eq.~(8) & $\lambda_{\rm h}$ &   1   &  $3.57\times 10^{-8}$  & &   & \\
\hline
B1 & $M_{\rm He}$ & $\lambda_{\rm b}$ &   1   &  $5.89\times 10^{-6}$ & 0.488 & 0.341 & 0.232 \\
B2 & $M_{\rm He}$ & $\lambda_{\rm g}$ &    1   &  $6.54\times 10^{-6}$  & 0.488 & 0.341 & 0.266 \\
\hline
C1 & $M_{\rm He}$ & $\lambda_{\rm h}$ &   1   &  $5.51\times 10^{-6}$ & 0.490 & 0.342 & 0.255 \\
C2 & $M_{\rm He}$ & $\lambda_{\rm h}$ &   $\propto q$   &  $1.74\times 10^{-6}$ & 0.502 & 0.350 & 0.389 \\

\hline
D1 & $M_{\rm CO}$ & $\lambda_{\rm h}$ & 1 &    $3.14\times 10^{-6}$ & 0.491 & 0.342 & 0.228 \\
D2 & $M_{\rm CO}$ & $\lambda_{\rm h}$ &  $\propto q$   &  $9.68\times 10^{-7}$ & 0.510 & 0.355 & 0.351 \\

\hline
\end{tabular}
\end{table}
\begin{table}

  \caption{The Kolmogorov-Smirnov statistic for the orbital period (upper) and
  effective temperature (lower) in each model. See Table~\ref{table2} for the values of
  $D_{\rm cr}(\alpha)$. \label{table3}}

   \begin{tabular}{ccccc}
  \hline Model &  Wind1/$\alpha_{\rm CE}=1$  &  Wind2/$\alpha_{\rm CE}=1$   & Wind1/$\alpha_{\rm CE}=0.2$ & Wind1/$\alpha_{\rm CE}=0.2$  \\
  \hline
  B1&0.274&0.316&0.355&0.304\\
  B1&0.086&0.108&0.110&0.372\\
  B2&0.315&0.322&0.303&0.405\\
  B2&0.107&0.095&0.519&0.633\\
  C1&0.263&0.290&0.342&0.315\\
  C1&0.129&0.090&0.113&0.215\\
  D1&0.290&0.324&0.335&0.311\\
  D1&0.122&0.119&0.146&0.166\\
  \hline

  \end{tabular}
\end{table}


\bsp	
\label{lastpage}
\end{document}